\documentclass[useAMS,usenatbib]{mn2e}
\usepackage{amssymb}
\usepackage{amsmath}
\usepackage{multirow}
\usepackage{epsfig}

\def\Journal#1#2#3#4{{#1} {\bf #2}, (#3) #4}
\def\cir#1{{\GCN} #1}
\def\rep#1{{\GCR} #1}

\def\etal{{\it et al.}}
\def\AA{\em A.\& A.}
\def\AIP{\em AIPC}

\def\APJ{\em ApJ.}
\def\APL{\em ApJ.Lett.}

\def\ASS{\em Astrophys.Space.Sci.}

\def\GCN{\em GCN Circ.}
\def\GCR{\em GCN Rep.}

\def\MRA{\em MNRAS}
\def\MRAl{\em MNRASL}
\def\NAS{New Astron.}
\def\NAT{\em Nature}
\def\NCA{{\em Nuovo Cimento} B}

\def\POF{\em Phys. of Fluids}
\def\PPC{\em Plasma Phys.Control.Fusion}
\def\PPL{\em Phys. Plasma}

\def\PRV{\em Phys. Rev.}

\def\PTR{{\em Phil.Trans.Roy.Soc.Lond.} A}

\def\SSR{\em Space Sci. Rev.}

\def\be{\begin{equation}}
\def\ee{\end{equation}}
\def\bea{\begin{eqnarray}}
\def\eea{\end{eqnarray}}
\def\bes{\begin{equation*}}
\def\ees{\end{equation*}}
\def\beas{\begin{eqnarray*}}
\def\eeas{\end{eqnarray*}}

\def\swift{{\it Swift }}

\title{A Systematic Description of Shocks in Gamma Ray Bursts: I. Formulation}

\author[H. Ziaeepour]{Houri~Ziaeepour
\thanks{Email: hz@mssl.ucl.ac.uk}\\ 
Mullard Space Science Laboratory, Holmbury St Mary, Dorking, 
Surrey RH5 6NT, UK}

\begin{document}
\date{Accepted $\ldots$; Received $\ldots$; in original form Dec. 2008}

\pagerange{\pageref{firstpage}--\pageref{lastpage}} \pubyear{2008}

\maketitle

\label{firstpage}

\begin{abstract}
Since the suggestion of relativistic shocks as the origin of gamma-ray 
bursts (GRBs) in early 90's, the mathematical formulation of this process has 
stayed at phenomenological level. One of the reasons for the slow development 
of theoretical works has been the simple power-law behaviour of the afterglows 
hours or days after the prompt gamma-ray emission. It was believed that they 
could be explained with these formulations. Nowadays with the launch of the 
\swift satellite and implementation of robotic ground follow-ups, gamma-ray 
bursts and their afterglow can be observed in multi-wavelength from a few 
tens of seconds after trigger onward. These observations have leaded to the 
discovery of features unexplainable by the simple formulation of the shocks 
and emission processes used up to now. Some of these features can be 
inherent to the nature and activities of the GRBs central engines which are 
not yet well understood. On the other hand {\bf\it devil is in details} and 
others may be explained with a more detailed 
formulation of these phenomena and without adhoc addition of new processes. 
Such a formulation is the goal of this work. We present a consistent 
formulation of the kinematic and dynamics of the collision between two 
spherical relativistic shells, their energy dissipation, and their 
coalescence. It can be applied to both internal and external 
shocks. Notably, we propose two phenomenological models for the evolution of 
the emitting region during the collision. One of these models is more suitable 
for the prompt/internal shocks and late external shocks, and the other for the 
afterglow/external collisions as well as the onset of internal shocks. 
We calculate a number of observables such as flux, lag between 
energy bands, and hardness ratios. One of our aims has been a formulation 
enough complex to include the essential processes, but enough simple such 
that the data can be directly compared with the theory to extract the value 
and evolution of physical quantities. To accomplish this goal, we also suggest 
a procedure for extracting parameters of the model from data. 
In a following paper we numerically calculate the evolution of some simulated 
models and compare their features with the properties of the observed 
gamma-ray bursts.
\end{abstract}


\section{Introduction} \label{sec:intro}
The \swift~\citep{swift} observations of more than 200 Gamma-Ray Bursts (GRBs) and 
their follow-ups have been a revolution in our knowledge and understanding of 
these elusive phenomena. The rapid slew of the \swift X-ray and UV/optical 
telescopes - respectively XRT~\citep{xrt} and UVOT~\citep{uvot} - as well as 
ground based robotic telescopes have permitted to observe GRBs and their 
afterglow in multi-wavelength from few tens of seconds after the prompt or 
precursor gamma-ray emission is detected by BAT~\citep{bat}, up to 
days after trigger. These observations show that the emission can be 
essentially divided to three regimes: 1) The prompt 
gamma-ray emission which can be very short, few tenth of milliseconds, or 
long, up to few hundred of seconds; 2) A tail emission in X-ray which is 
observed for more than $90\%$ of bursts. For some bursts this tail is also 
detected as a soft faint continuum in gamma-ray. In about $40\%$ of bursts 
this early emission has been detected in optical and infrared too. In this 
regime for many bursts flares have been observed mainly in X-ray. Sometimes 
the counterpart of flares have been also observed in gamma-ray and/or 
optical/IR. In many bursts the early steep slope of the X-ray emission at the 
beginning of this regime becomes much shallower and somehow harder at 
the end; 3) The late emission can be considered as the epoch after the break 
of shallow regime in which the emission is usually a continuum and no or 
little flaring activity is observed (but there are exceptions such as 
GRB 070110~\citep{grb070110} and GRB 081028~\citep{grb081028} which had 
bright late flares). The duration and relative fluxes of these regimes 
can vary significantly between GRBs.

In one hand, it seems that the idea of synchrotron emission from accelerated 
particles in a relativistic shock as the origin of the prompt 
emission~\citep{intext,intext1} is essentially correct. On the other hand, 
the early observations of what is usually called {\it the afterglow} - 
the emission in lower energy bands usually observed from $\lesssim 100$~sec 
after trigger onward - have 
been the source of surprises and raised a number of questions about many 
issues: the activity~\citep{longact} and the nature of the 
engine~\citep{progenitor,progenitor1,progenitorshort}, the concept of 
prompt/internal-afterglow/external 
shocks~\citep{promptext}, the efficiency of energy transfer from the 
outflow -{\it the fireball} - to synchrotron radiation~\citep{enereff}, the 
collimation and jet break~\citep{jetbreakex}, the behaviour of X-ray and 
optical light curves~\citep{optag}, etc. Many of predictions such as the 
existence of a significant high latitude emission with a strict relation 
between the light curve time evolution slope and the spectrum index, and an 
achromatic jet break have not been observed. Moreover, the origin of 
unexpected behaviours such as a very steep decline in low energy bands after 
the prompt~\citep{tailhighlat} and a very shallow regime which lasts for 
thousands of seconds are not well understood. Other unexpected observations 
are the existence of the chromatic multiple breaks in the X-ray light curves, 
flares in X-ray and optical bands hundred of seconds after the prompt even 
in some short bursts (ex: GRB 060313~\citep{grb060313,grb060313-1}, 
GRB 070724A~\citep{grb070724a}, a tail emission in short bursts
(ex: GRB070714B~\citep{grb070714b,grb070714b1}, GRB080426~\citep{grb080426}, 
GRB 080503~\citep{grb080503}), and very short, hard, and high amplitude 
spikes in long bursts that could lead to the classification of the burst as 
short if the instrument was not enough sensitive to detect the rest of the 
prompt emission (ex: GRB060614~\citep{grb060614,swiftgrb060614}, 
GRB061006~\citep{grb061006}). This makes the classification of 
bursts as short and long much more ambiguous~\citep{grbnewclass}.

One conclusion that has been made from these observations is that the central 
engine can be active for up to thousands of seconds after the prompt 
emission~\citep{longact}. But the nature of the fireball and its source of 
energy is not yet well understood, and we can not yet verify this 
interpretation or relate 
it to any specific process in the engine. It seems however that 
whatever the origin of the fireball, it must be baryon dominated otherwise 
it could not make long term effects correlated to the prompt emission. In 
this case, the internal and external shock models as the origin of the 
prompt and afterglow are good candidates. Nonetheless, the lack of a simple 
explanation for the observed complexities has encouraged authors to 
consider other possibilities, for instance associating both the prompt 
gamma-ray and the afterglows to external shocks and fast variations to abrupt 
density variation of the surrounding material~\citep{promptext}. However, 
it has been shown that in such models it is not possible to explain the fast 
variations of the prompt even in presence of a bubbly environment or 
pulse-like density change~\citep{promptext,promptext1}. 

Here we suggest that at least some of the features of early afterglows can be 
related to a complex shock physics and/or features in the fireball/jet. In 
fact, 
simulations of the acceleration of electrons and positrons by the first and 
second Fermi processes show that the evolution of electric and magnetic 
fields as well as the energy distribution of accelerated particles are quite 
complex~\citep{fermiacc,fermiacc1,fermiacc2}. Plasma instabilities lead to 
the formation of coherent electric and magnetic fields and acceleration of 
particles~\citep{weibel,weibel1,weibeltemp,instabparal}. Their time 
evolution in relativistic shocks can significantly affect the behaviour of 
the prompt and the afterglow of GRBs. If the number density of particles in 
the ejecta is significant and the shock is collisional, the state of matter 
in the jet can be also an important factor in determining the behaviour of 
the fields, and thereby the synchrotron emission by accelerated electrons 
and positrons. Many aspects of these processes are not well understood, 
however realistic interpretations of observations should consider these 
complexities at least phenomenologically. For instance, the simple 
distributions such as a power-law distribution for Lorentz factor of 
electrons, or a constant magnetic field for the whole duration of prompt 
and afterglow can be quite unrealistic. Ideally, these quantities should 
come from the simulation of Fermi processes and plasma instabilities such as 
Weibel instability~\citep{weibel,weibel1} that produces the coherent 
transverse magnetic field. However, these phenomena 
are complex and their simulations are very time and CPU consuming. For these 
reasons they can not yet explore the parameter space of the phenomena 
and are mostly useful for demonstrating the concepts and how they work. 
Therefore we are obliged to use simple analytical approximations for 
quantities related to the physics of relativistic shocks. In this situation 
a compromise between complex non-analytical expressions and too simplistic and 
too simplified but unrealistic analytical behaviour of the physical 
quantities can be the consideration of intervals in which a simple analytical 
function can be a good approximation. Then, by adding together these 
intervals - regimes - one can reconstruct the entire evolution of a burst 
and its afterglow.

Even with a simplified presentation of the physical processes one would not 
be able to explain GRB data without a model including both microphysics and 
dynamics of the fireball. The majority of works on the modelling of shocks 
and synchrotron emissions either deal with the 
emission~\citep{emission,emission0,emission1,revshock,revshocksimul} or with 
the kinematics of the shock~\citep{kinematic,kinematic1,kinematic2}, or both 
but in a phenomenological way~\citep{kinematic3}. Few 
works~\citep{shocksynch,shocksynch1} have tried to include both these aspects 
in a consistent model, but either they have not been very successful - their 
predictions specially for quantities such as lags in different bands were far 
from observed values and additional parametrization was necessary - or the 
formulation is too abstract to be compared directly with 
data~\citep{grbboltz,grbboltz1}. 

With these issues in mind, in Sec.\ref{sec:model} we present a simplified 
shock model that includes both the kinematics of the ejecta and the dynamics 
of the synchrotron emission. The microphysics is included by the means of a 
simple parametrization. We calculate a number of observables such as flux, 
hardness ratios, and lags between different energy bands.
In this paper and paper II in which we simulate part of prompt and afterglow 
regimes of GRBs in some time intervals, we show that this model can explain 
many aspects of bursts 
as long as we divide the data to separate regimes. The reason is that the 
simple parametrization of microphysics in this model can be valid at most in 
a limited time interval. Each regime should be separately compared to 
analytical and numerical results for extracting the parameters. The results 
will show how parameters that are considered as constant in this model evolve 
during the lifetime of the burst. This is the best we can do until a better 
understanding of relativistic shock models and Fermi processes become 
available. If the model and the estimation of its parameters for each 
regime is sufficiently correct, adding them together should give us an 
overall consistent picture of characteristics of the burst, its 
afterglow, and its surrounding material. Apriori this knowledge should help 
to better understand the engine activities and eventually its nature and 
classification.

The model presented here depends on a large number of parameters and we need 
an extraction procedure permitting to extract as much as possible information 
about the physical properties of the shock from the available data. In 
Sec.\ref{sec:extract} we explain how in the frame work of this model one can 
extract various quantities from data. Evidently the success of the 
modelling strongly depends on availability and quality of simultaneous 
multi-wavelength observations.

The \swift observations show that during the first few hundred seconds after 
the trigger there is usually a very close relation between the prompt 
gamma-ray emission and the emission in lower energy bands~\citep{earlyemi,
earlyemi1}, 
therefore, most probably they have a common origin, presumably internal 
shocks. However, historically and even in the present literature (and 
sometimes believes) any emission after the prompt gamma-ray is called 
{\it the afterglow} - meaning due to a shock with ISM or surrounding material, 
presumably external shocks. Therefore, for clarity of context here we define 
{\it the afterglow} as the emission in any energy band and at any time after 
the main prompt peaks regardless of its origin. If by {\it afterglow} we mean 
the external shocks, this is mentioned explicitly in the text.

We finish this paper with some outlines and two appendices containing the 
details of calculation of the dynamics and flux for power-law distribution of 
electrons Lorentz factor.

\section{Shock Model} \label{sec:model}
In this section we first give a sketch description of a relativistic collision 
between two shells of material and processes which produce gamma-ray and 
other radiations. Then, we discuss a simplified mathematical formulation of 
the evolution of the shock and synchrotron emission. By restricting the model 
to a thin layer and to the early times after beginning of the collision, 
we can analytically calculate various observables.

\subsection{Qualitative description of a relativistic shock and its 
simplified model} \label{sec:qualdesc}
We begin this section by a pictorial description of present believes about the 
origin of GRBs and their afterglow. A central source - supernova, collapsar, 
collision of two compact objects, etc. - ejects highly relativistic cold 
baryon dominated material as a spherical, jet or torus-like shells 
call {\it the fireball}. Collision between faster later ejected shells with 
slower earlier ejected ones produces what is called the prompt 
emission~\citep{intext}. 

Apriori, there is no reason why faster shells 
should be ejected later. One way of explaining this paradigm is the 
deceleration of the front shells~\citep{shelldecel} by surrounding material 
which are observed around massive objects such as Wolf-Rayet (WR) 
stars~\citep{ismwrstar}. It is possible that this initial deceleration is the 
source of a weak emission which has been seen before the main spike in many 
bursts. 
A relatively weak and soft precursor spike has been also observed in some 
bursts and can be related to this deceleration~\citep{precursdecel}. 
Although other origins such as jet-star 
interaction~\citep{precursjetstar,precursjetstar1} and fallback to the 
collapsar~\citep{precursfall} are also suggested to explain precursors, 
deceleration of the initial shell seems to be a more natural explanation and 
does 
not need any fine tuning of the progenitor models and their parameters. In 
contrast, jet-star interaction scenario can not explain large time lag - few 
100 of seconds between precursor and the main spike in some bursts e.g. 
GRB 050820A~\citep{grb050820a,grb050820a1}, 
GRB 060124~\citep{grb060124,grb060124-1}, and GRB 070721B~\citep{grb070721b}. 
In the fallback scenario a weak 
jet is produced during the formation of a transient neutron star which later 
collapses to a black hole. The main jet in this scenario is produced by the 
accretion of the material from a disk to the black hole. In this case, the 
lag depends on the lifetime of the proto-neutron star and the rate of the 
accretion from the surrounding disk. Although these parameters can be tuned 
to explain the lag, apriori much longer lags should be also possible but 
never observed. In the deceleration scenario the maximum possible lag is 
the duration of the central engine main activity - few hundred of seconds 
according to the observations of the main flares in X-ray, and is consistent 
with all the observations. The UV emission from the precursor should 
ionize the unshocked material in front of the first shell and therefore there 
would be little additional absorption of soft X-ray later~\citep{agreion}. 
In fact in GRB 060124~\citep{grb060124,grb060124-1} which had long lag 
between the precursor and the main peak, a slight increase in $N_H$ column 
density at late time with respect to the initial density has been detected.
In Paper II we argue that another possible origin of the precursor is the 
temporary dynamical reduction of the emission in the early stage of the 
shock that make the burst unobservable for a short time. Then, with the 
progress of coalescence of the shells the emission resumes and is observed 
as the main peak.

\begin{figure*}
\begin{center}
\begin{tabular}{cc}
a)\includegraphics[height=5cm]{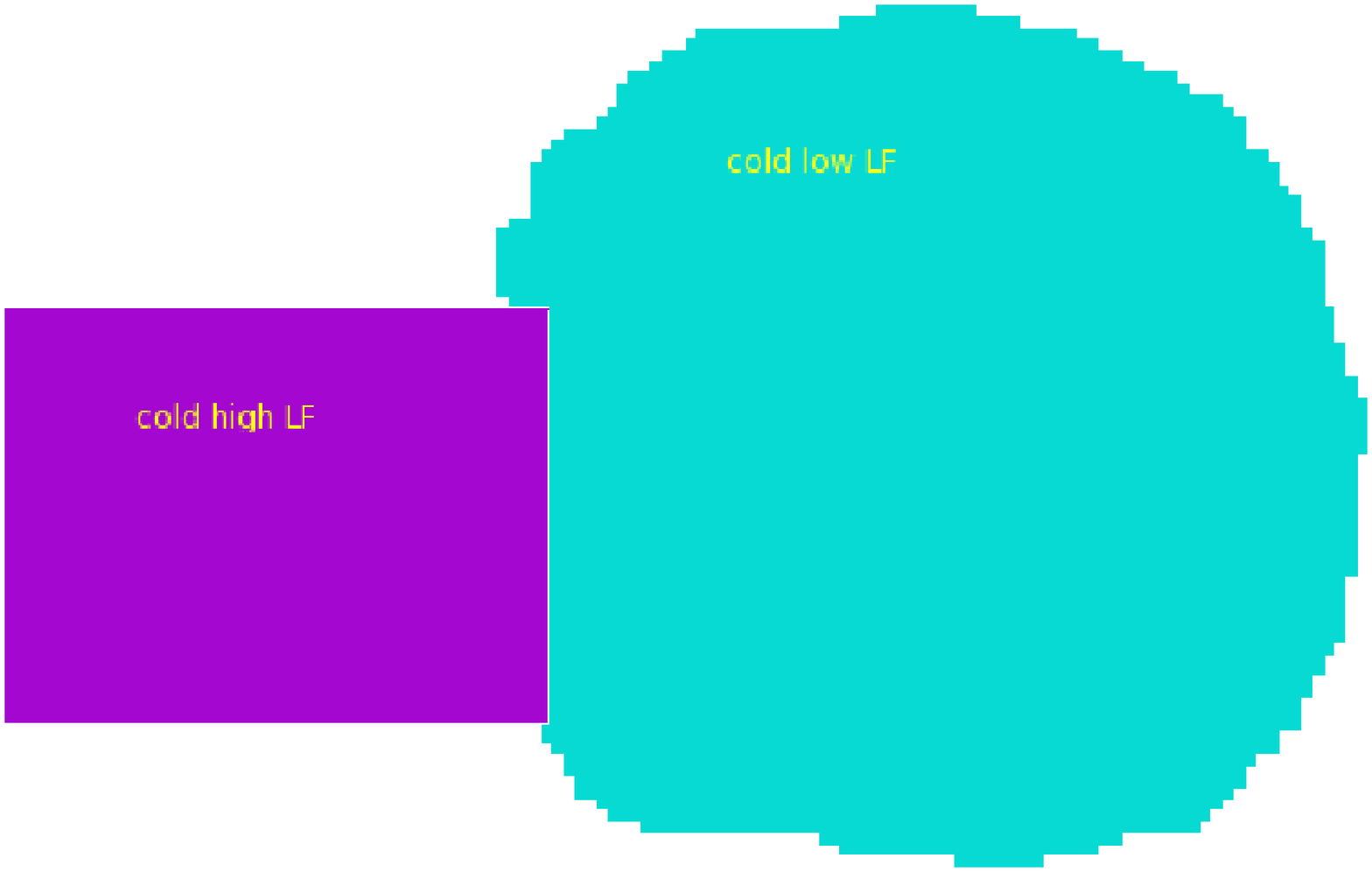} &
b)\includegraphics[height=5cm]{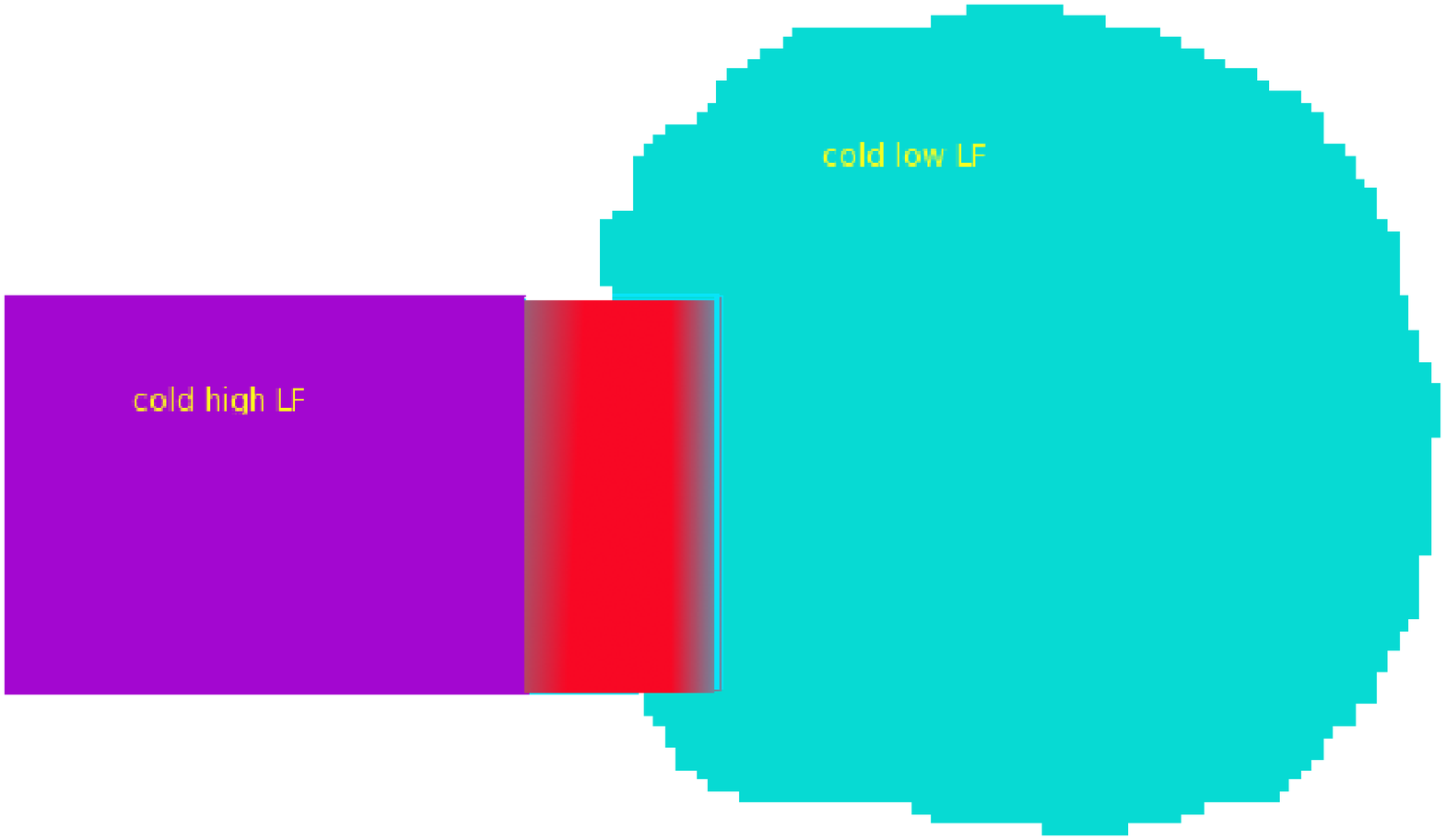} \\
c)\includegraphics[height=5cm]{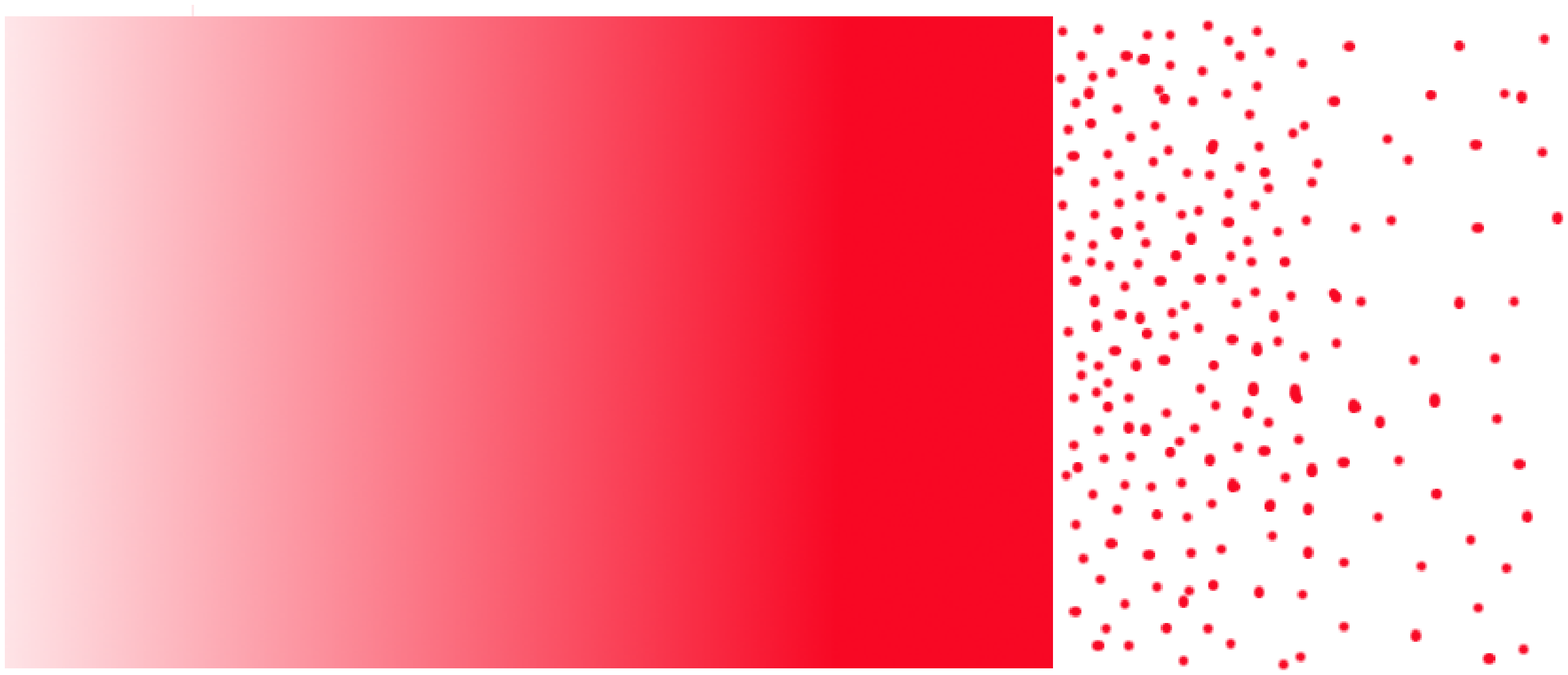} &
d)\includegraphics[height=5cm]{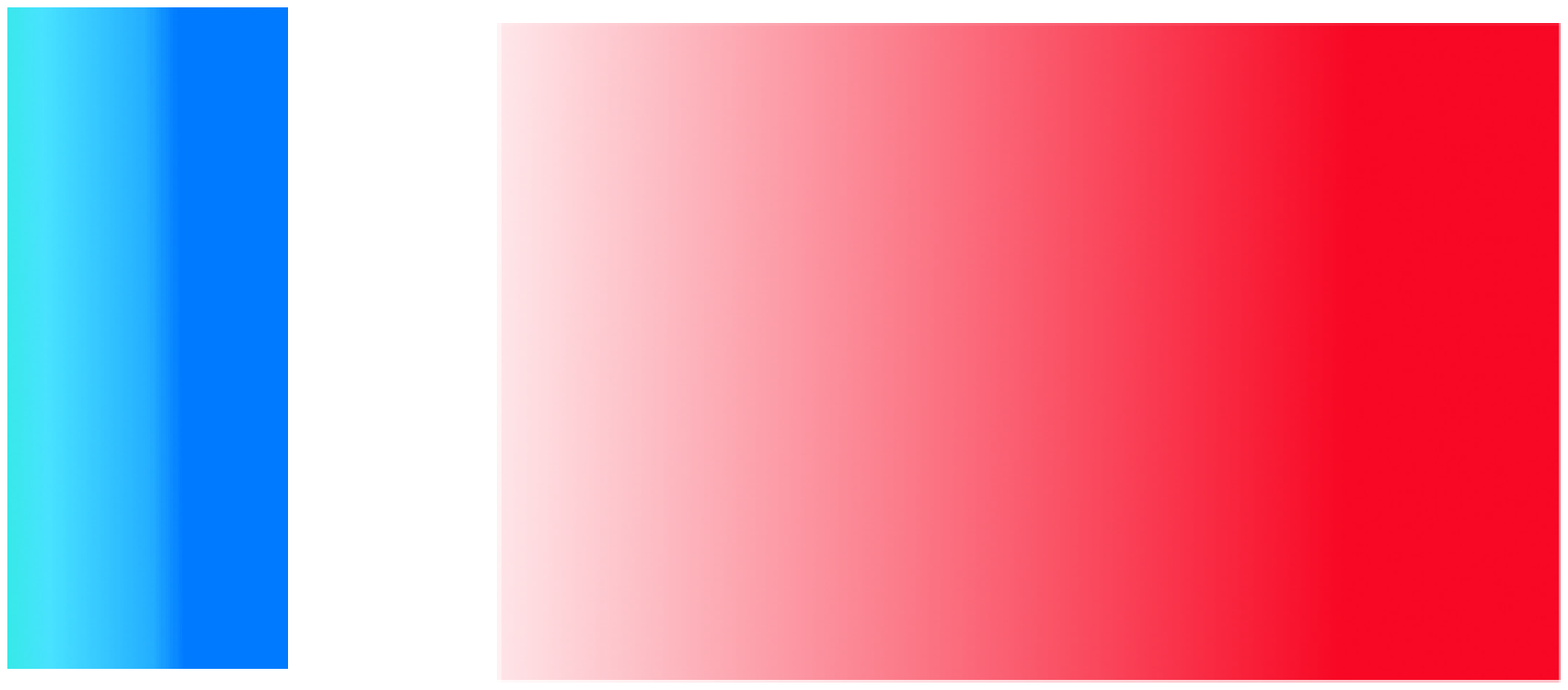}
\end{tabular}
\caption{Sketch of a relativistic shock: a) Beginning of the shock: A high 
Lorentz factor cold shell (violet) moving from left to right collides with a 
slower shell or ISM (blue). b) During the passage of the shells through each 
other, at the shock front an {\it active region} is formed where large 
electric and orthogonal magnetic fields are induced by plasma instabilities 
and charged particles are accelerated. They lose part of their kinetic energy 
as a synchrotron emission. The concept of active (emitting) region is not new 
and is commonly used in the context of the formation of cosmic rays in 
relativistic and non-relativistic shocks by Fermi 
processes~\citep{crfermi,crfermi1}. A fraction of 
these highly accelerated particles escape to the downstream. By contrast, 
particles that have lost their energy move toward upstream. This process 
extends the active region but reduces the gradient of quantities 
such as density, thus gradually weakens the shock. c) and d) show two 
possible outcome of the collision: total coalescence of the shells (radiative 
collision) c) and when after the passage of the fast shell a left-over slow 
shell is formed behind. In these figures colour gradient presents the 
velocity and density distributions: darker colours correspond to higher 
velocity/density.
} \label{fig:shochcaricature}
\end{center}
\end{figure*}

When two high density shells collide, in the most general case they partially 
coalesce. Then, the most energetic particles get ahead of the rest in the 
downstream, and at the end of the collision the configuration includes again 
two shells. The back shell consists of slower particles that have lost their 
kinetic energy (shocked particles). The front shell is the remnant of faster 
unscattered particles in the shells. In practice we expect that the kinetic 
energy difference be continuous and slower particles become an expanding 
tail behind a faster and most probably denser head. In few bursts such 
as GRB060607A~\citep{grb060607a,grb060607a1} and GRB 070107~\citep{grb070107} 
it seems that we are seeing the separation between these components in the 
X-ray afterglow. Fig.\ref{fig:shochcaricature} shows a sketch of the shock 
processes.

In the simplest case the shock between two shells is radiative. This means 
that for an observer in the rest frame of the fast shell the kinetic energy 
of the falling particles from the other shell is immediately radiated and 
particles come to rest and join the shell. For a far observer at rest with 
respect to the engine the difference between the kinetic energy of the two 
shells is partially radiated and partially transferred to the particles of 
the slower shell. The fast shell is decelerated until the totality of the 
slow shell is swept. 

Not all the particles in the fast shell are decelerated with the 
same rate. Therefore after the coalescence of the shells there is a gradient 
of Lorentz factor from the shell front head with highest $\Gamma$ to the back 
tail with lowest $\Gamma$. In the macroscopic treatment of the shock 
processes it is usually assumed that during the collision two distinguishable 
shocked layers and corresponding discontinuities are formed in each side of 
the boundary between the shells. They are called forward and reverse shocks 
according to their evolution direction with respect to the initial 
discontinuity~\citep {emission0,revshock,revshocksimul}. According to these 
models and 
depending on the density difference of the shells and their relative Lorentz 
factor, the induced electric and magnetic fields and thereby the distribution 
of accelerated electrons and their synchrotron emission in these shocked 
regions can be very different. Notably the reveres shock is expected to emit 
mostly in optical wavelength with a relatively large lag with respect to the 
prompt gamma-ray as the emission must traverse the width of both shocked 
layers. Therefore it should appear as an optical flash asynchronous from 
gamma-ray peaks~\citep{opticalflash}.

On the other hand, if shells have similar 
densities and a small relative Lorentz factor, the scattering of particles 
in the two sides of the boundary quickly homogenizes the shocked regions, 
and therefore one can consider a single shocked zone with a relatively 
slow gradient in density and fields. If one of the shells has a density much 
larger than the other, the width of its shocked layer would be very small 
with respect to the shocked region in the other shell and again the 
assumption of a single shocked layer is a good approximation for internal 
shocks. This schematic view corresponds well to what is observed in the 
simulations of Fermi acceleration in the ultra-relativistic shocks (see Fig.1 
of ~\cite{fermiaccspec} and ~\cite{shockmag}). 

For a far observer who only detects photons from the synchrotron emission of 
the shocked material it is very difficult to distinguish between photons 
coming from distinct regions unless they are well separated in time and in 
energy band. The lack of a clear evidence for a reverse shock emission in the 
\swift bursts, specially during the prompt emission, means that the reverse 
shock in the prompt GRB emission is weak and the assumption of just one 
shocked region is a good approximation. Most of the synchrotron emission is 
expected to be emitted by charged particles in the shocked region. But as 
electric and magnetic fields, and accelerated particles are not really 
confined in this region and penetrate to a larger area, the region that 
emit radiation can be much extended than shocked region. We call this emitting 
zone {\it the active region}.

Evidently, in a real situation multiple shells are ejected in a short time 
interval. In this case both pair collisions between late shells and 
collision-coalescence of the late shells with the main ejecta - 
{\it prompt shell} - is possible. These events happen at different points 
of space-time, but their synchrotron emission can arrive to the observer 
separately, partially overlapping, or simultaneously. Therefore it is not 
always possible to distinguish separate collisions and their characteristics. 
To this complexity one must also add the variation of quantities such as 
density and Lorentz factor in a single shell. They increase the variability 
of observed emissions. On the other hand, overlapping emissions make the 
comparison of data with the models based on a simple parametrization of the 
physical properties of a shell ambiguous. Despite these complexities one 
should be able to consider peaks as separate collisions and find an 
{\it effective} set of parameters to model each one separately.

\subsection{Shock evolution} \label{sec:shockevol}
A shock is defined mathematically as a discontinuity in the density 
distribution of a flow. It should satisfy at each point of the space-time the 
total energy-momentum and current/particle flux conservation 
equation~\citep{relshock}:
\bea
T^{\mu\nu}_{;\mu} &=& 0 \label{enermom} \\
\sum_i (\rho_i u_i^{\mu})_{;\mu} &=& 0 \label{current}
\eea
where $T^{\mu\nu}$ is the total energy-momentum tensor; $\rho$ is density 
and $u^{\mu}$ velocity vector; the index $i$ indicates species of 
particles/fluids with a conserved number. They also include the interactions 
between these particles/fluids. 
When there is no interaction, these equations must be satisfied separately 
for each species. In particular, at the shock front where these quantities 
are discontinuous the conservation equations take the form of jump 
conditions across the discontinuity surface:
\bea
[T^{\mu\nu}]~\Sigma_{\mu} &=& 0 \label{jumpener}
\eea
\bea
\sum_i [{\rho}_i u_i^{\mu}]~\Sigma_{\mu} &=& 0 \label{jumpcurrent}
\eea
The symbol $[~]$ means the difference of the quantities between square 
brackets on the two sides of the shock front. Solutions of these equations 
determine the evolution of the shock front. As for the state of the shocked 
material behind 
the discontinuity, when there is no energy dissipation jump conditions can be 
used to obtain a self-similar solution~\citep{kinematic}. In presence of 
energy injection or 
dissipation however the self similarity solutions are only approximations and 
in general an exact self similar solution does not exist. Moreover, energy 
dissipation by synchrotron emission changes the form of the conservation 
equations, i.e. it can not be treated as an inhomogeneous term in the 
differential equations describing the dynamics. Thus, it is not 
possible to extend the solutions of the non-dissipative case to this case 
even as an approximation. We will discuss this issue in details later in this 
section. Therefore in a dissipative shock the jump conditions are applied only 
at the initial time when a shell meets another shell or the ISM, and they 
should be considered as boundary conditions when conservation equations 
(\ref{enermom}) and (\ref{current}) are solved.

Intuitively we expect that with time the discontinuous distributions of 
quantities in the shock front i.e the initial jump conditions change to 
continuous distributions which at far downstream distances should 
asymptotically approach to the slow shell values, and at far distances in the 
upstream direction to the characteristics of the fast shell - see 
Fig.\ref{fig:shochcaricature} for a schematic illustration. In this transient 
region instabilities form electric and transversal magnetic fields. They 
accelerates electrons which subsequently lose their energy by synchrotron 
and inverse Compton emission. A far observer receives the signature of the 
shock mainly through the detection of these radiations as $\gamma$-ray burst 
or X-ray 
flare. Therefore finding the evolution of physics of this region is the main 
purpose for solving conservation equations.

A full solution of equations (\ref{enermom}) and (\ref{current}) with the 
initial conditions (\ref{jumpener}) and (\ref{jumpcurrent}) apriori include 
all the necessary information about the physical processes and their 
evolution. However, the complexity of the problem can not permit to solve 
them analytically. On the other hand, numerical simulations are 
both complex and time consuming, and it would be very difficult to cover the 
full parameter space and obtain results that can be compared with the 
observations. At present, simulations are only able to demonstrate the 
validity of ideas about processes involved and the role of the 
instabilities in the formation of coherent fields~\citep{fermiaccspec,shockmag}.

Here we consider an intermediate strategy. We assume a spherical\footnote{Most 
of the formulation presented here is also applicable to a beamed ejecta/jet 
if the transverse dispersion of matter in the jet is negligible with 
respect to the boost in the radial direction. We consider the effect of 
beaming in Sec.\ref{sec:break}} thin and optically transparent active region. 
Its average distance from the central source is $r(t)$ and its thickness 
$\Delta r(t)$. We neglect the variation of quantities inside the active 
region and consider the average value through the region. This means that the 
evolution depends only on $r(t)$ the mean distance of the active region from 
central engine. As for the energy 
dissipation, we assume a radiative shock i.e. for an observer at rest with 
respect to the active region the incoming particles lose their energy through 
synchrotron radiation and join this region. Despite possibility of 
large contribution from Compton cooling of electrons in the GRBs 
prompt~\citep{comptonself,comptontev}, X-ray flares~\citep{comptonreverse}, 
and 
afterglow~\citep{comptonag}, there is no strong evidence of this process in 
the \swift data. Cases such as the early flare of 
GRB 050406~\citep{grb050406} that was once attributed to Compton cooling of a 
reverse shock is now understood to be related to the late activity of the 
central engine like all other observed flares and have the same origin as 
the gamma-ray. It is also suggested that the anomalous behaviour of 
GRB 060218/SN 2006aj~\citep{grb060218} is due to inverse Compton 
cooling~\citep{grb060218comp}, but this is an exception. Therefore here we 
only consider synchrotron radiation in a relativistic shock as the source 
of emission. 

From now on we consider the active region as an isolated shell of baryonic 
material. The effect of ISM/slow shell appears as an incoming flux, and the 
effect of diffusion of shocked material into the upstream is included in the 
evolution of the thickness of the active region and other physical parameters 
such as density variation with time. The initial value of the density and 
Lorentz 
factor correspond to the values in the fast shell i.e. the shock front at 
the time of encounter between the two shells (internal shock) or fast shell 
and the ISM (external shock). We also assume that thermal energy and pressure 
are negligible with respect to the relativistic boost. This assumption of 
cold matter is specially justified for the prompt emission because the 
temperature of ejected material from the progenitor is expected to be at 
most a few hundred MeV, where the kinetic energy of relativistic baryons 
must be of order of few GeV or larger. This approximation is not probably 
suitable for late interaction of the shells when they have lost most of their 
kinetic energy and probably turbulence and scattering have transferred part 
of this energy to thermal.

With these approximations we write the energy-momentum conservation equations 
for the active region in the rest frame of the slow shell. The reason for 
choosing this frame is that the same formulation can be applied to external 
shocks where the slow shell is the ISM or surrounding material around the 
engine. The latter is 
considered to be at rest with respect to the observer (after correction for 
the expansion of the Universe)\footnote{Through out this work frame-dependent 
quantities with a prime are measured with respect to the rest frame of the 
slow shell and without prime with respect to a far observer at the redshift 
of the central engine. Parameters used for parametrization do not have a prime 
even when parametrization is in the slow shell frame.}. Note that for the 
consistency and conservation of energy and momentum we have to integrate 
equations (\ref{enermom}) and (\ref{current}) along the active region. As we 
only consider the average value of quantities, the integration is 
trivial\footnote{Note also that apriori we should consider an angular term 
presenting the collimation of the radial element. But for an infinitesimal 
element the angular dependence is negligible and we do not add it to this 
formulation.}:
\bea
&&\frac {d(r'^2 n' \Delta r' \gamma')}{dr} = \gamma' \biggl (r'^2 
\frac{d(n'\Delta r')}{dr'} + 2r' (n'\Delta r')\biggr ) + \nonumber \\
&&\quad\quad r'^2 (n'\Delta r') \frac{d\gamma'}{dr'} = 
n'_0(r) r'^2 - \frac{dE'_{sy}}{4\pi m c^2dr'} \label {enercons} \\
&& \frac {d(r'^2 n' \Delta r' \gamma' \beta')}{dr'} = \beta' \gamma' (r'^2 
\frac{d(n'\Delta r')}{dr'} + 2r' (n'\Delta r')) + \nonumber \\
&&\quad\quad r'^2 (n'\Delta r') \frac {d(\beta' \gamma')}{dr'} = 
- \frac{dE'_{sy}}{4\pi m c^2dr'} \label {momcons}
\eea
where $r'$ is the distance from the central engine, $n'$ is the baryon number 
density of the fast shell measured in the slow shell frame, $n'_0$ is the 
baryon number density of the slow shell in its rest frame and in general it 
can depend on $r'$. Here we assume that $n'_0(r') = N_0 (r'/r'_0)^{-\kappa}$. 
For ISM $\kappa = 0$, i.e. no radial dependence. For a wind surrounding the 
central engine $\kappa = 2$ is usually assumed~\citep{windexpo}. For a thin 
shell or jet expanding adiabatically also $\kappa = 2$ if we neglect 
the transverse 
expansion in the case of a jet (collimated ejecta). But for the collision 
between two thin shells if the duration of the collision is much smaller than 
$r'_0/c$ we can neglect the density change due to expansion during the 
collision, and assume $\kappa = 0$. $\Delta r'$ is the 
thickness of the shocked synchrotron emitting region, $\gamma'$ is the 
Lorentz factor of the fast shell with respect to the slow shell, 
$\beta' = \sqrt {\gamma'^2 - 1} / \gamma'$, $m = m_p+m_e \approx m_p$, 
$E'_{sy}$ is the total emitted energy, and $c$ is the speed of light. 
The evolution of the radius with time is:
\be
r' (t') - r' (t'_0) = c \int_{t'_0}^{t'} \beta' (t'') dt'' \label {revol}
\ee
where the initial time $t'_0$ is considered to be the beginning of the 
collision. 

Apriori we should also consider a conservation equation for the baryon and 
lepton numbers. However, with approximations explained above the thickness 
of the active region is a parameter which is added by hand. The simplest 
assumption is that the active region is formed only from particles that  
fall from the slow shell to the shock front and they stay there. 
In this case the baryon number conservation is simply:
\be
\frac{d(n'\Delta r')}{dr'} = n'_0 \label{numcon}
\ee
On the other hand the assumption of particles staying for ever in the active 
region seems quite unphysical because scattering, acceleration, and 
dissipation move shocked particles to both upstream and down stream and 
gradually many of particles are in a region where instabilities are too 
weak to make the electric and magnetic fields necessary for the acceleration 
and synchrotron emission. Therefore in this approximation the active region 
can not be considered as an completely isolated system with full 
conservation laws applied to it. The consequence of this is that we can not 
determine $\Delta r$ from first principles. This point can be also interpreted 
as the manifestation of the fact that there 
is not an abrupt termination of the active zone, and therefore there is no 
conservation for the artificial boundary we have added by hand. In fact it is 
equally valid if we consider a conserved number of particles and the 
corresponding conservation equation but give up energy or momentum 
conservation. In this definition the active region follows 
{\it active particles} (in an Hamiltonian formulation sense). However, 
the only measurable quantity for a far observer is the energy dissipated as 
radiation. Therefore, it is more useful to define the active region based on 
the energy-momentum conservation and leave the number of these particles as a 
free parameter. Note 
also that in the left hand side of (\ref{enercons}) and (\ref{momcons}), as 
well as in the synchrotron term under some conditions (see below), 
$\Delta r'$ always appears as $n' \Delta r'$ i.e. the column density. It is 
more relevant for the observations and does not depend on the way we define 
the active region.

We can formally integrate equation (\ref{enercons}) by replacing the 
synchrotron term in the left hand side of (\ref{momcons}), and determine 
the column density of the active region\footnote{We remind that column 
density and total power are scalars and therefore their value is frame 
independent}:
\bea
&& n' \Delta r' = \nonumber \\
&& \hspace{-0.2cm} \frac {N_0 {r'}_0^3 ((\frac {r'}{{r'}_0})^
{3-\kappa} - 1) + (3-\kappa) {r'}_0^2 n'({r'}_0)\Delta r' ({r'}_0)\gamma'_0 
(1-\beta'_0)}{(3-\kappa) r'^2 \gamma' (1-\beta')} \nonumber \\
\label{columndenssol}
\eea
where $\gamma'_0 \equiv \gamma'(r'_0)$ and $\beta'_0$ is the corresponding 
$\beta'$. This solution depends on the evolution of $\gamma'$ which can be 
obtained by solving (\ref{momcons}), but we first remind the dependence of 
the synchrotron term on the microphysics of the shock. The power of 
synchrotron radiation emitted by the active shell is:
\bea
P'= \frac {dE'_{sy}}{dt'} = c\beta' \frac {dE'_{sy}}{dr'} & = & \frac {16\pi}{3} 
r'^2 \Delta r' \sigma_T c \gamma'^2 \frac{B'^2}{8\pi} \nonumber \\
& & \int n'_e (\gamma_e) \gamma_e^2 d\gamma_e \label {synchpower}
\eea
where $n'_e$ is the number density of accelerated charged leptons - electrons 
and possibly positrons - with a Lorentz factor $\gamma_e \gg \gamma'$ in the 
active region (shock) frame, $B'$ is the magnetic field - $B'^2/8\pi$ is the 
magnetic energy density in the active region frame, and $\sigma_T$ is 
Thompson cross-section. We define the normalization of electron distribution 
as the following:
\bea
\int_{\gamma_m}^{\infty} n'_e (\gamma_e) d\gamma_e & = & n'_a \label {enum} \\
\int_{\gamma_m}^{\infty} \gamma_e n'_e (\gamma_e) d\gamma_e & = & 
\frac {\gamma'^2 m_p n'_0 \epsilon_e}{m_e} \label {eener}
\eea
where $n'_a$ is the number density of accelerated charged leptons and 
$\epsilon_e$ is the fraction of the kinetic energy of the falling baryons 
transferred to the accelerated leptons in the active region frame. 

In GRB/relativistic shock literature it is usually assumed that 
$n'_a \approx \gamma' n'_0$, i.e. only falling leptons are 
accelerated. However, the validity of this assumption is not certain because 
once the electric and magnetic fields are produced by the instabilities, all 
the charged leptons are accelerated. If the initial number density of charged 
leptons in the rest frame of the shells is similar to the slow shell, then 
as the flux of falling 
leptons is enhanced by a factor of $\gamma'$, the density of local leptons can 
be neglected. However, in prompt collision one expects that the relative 
Lorentz factor of the shells be ${\mathcal O}(1)$. In this case the local density of leptons is 
not negligible and $n'_a \approx (n'_{te} + n'_{tp})$, with 
$n'_{te}$ and $n'_{tp}$ respectively the total number density of electrons 
and positrons. For a neutral matter with negligible positrons content 
$n'_a \approx n'_{te}$.

Motivated by the power-law distribution of accelerated charged particles in 
other astronomical shocks, e.g. supernovae and cosmic rays, it is usually 
assumed that the distribution of accelerated electrons responsible for the 
GRB prompt and afterglow emission is a power-law:
\bea
n'_e (\gamma_e) & = & N_e \biggl (\frac {\gamma_e}{\gamma_m}\biggr)^{-(p+1)} 
\mbox{for } \gamma_e \geqslant \gamma_m \label{nedistpow} \\
N_e & = & \frac {n'_a p}{\gamma_m} = \frac{p^2 m_e {n'}_a^2}{(p-1) \epsilon_e 
\gamma'^2 m_p n'_0} \label{neamp} \\
\gamma_m & = & \frac {(p-1)\epsilon_e \gamma'^2 m_p n'_0}{p m_e n'_a} 
\label{gammam}
\eea
Recent simulations of particle acceleration by Fermi process in the 
relativistic shocks show~\citep{fermiaccspec} that $n'_e (\gamma_e)$ is best 
fitted by a 2D Maxwellian distribution plus a power-law with an exponential 
cutoff:
\bea
n'_e (\gamma_e) & = & C_1 \gamma_e \exp(-\gamma_e/\gamma_1) + \nonumber \\
& & C_2 \gamma_e^{-\delta} \text{min}[1,\exp(-(\gamma_e - \gamma_i) / 
\gamma_{cut})] \label {fermispec}
\eea
where $C_2 = 0$ for $\gamma_e$ less than a minimum value $\gamma_{min}$. 
The typical values of parameters obtained from the fit to simulations for an 
initial Lorentz factor of $\gamma_0 = 15$ are: $\gamma_{min} = 40$, 
$\gamma_1 = 6$, $\gamma_i = 300$, $\gamma_{cut} = 100 $, and 
$\delta = 2.5$~\citep{fermiaccspec}. Implementation of this distribution makes 
the model presented here significantly more complex. In the range of 
energies relevant to the prompt and early afterglow emission of GRBs, the 
first term in (\ref {fermispec}) is negligible and for $\gamma_e > \gamma_i$ 
the distribution has the form of a power-law with exponential 
cutoff. Conservation conditions similar to (\ref{neamp}) and (\ref{gammam}) 
for this distribution lead to the following relations:
\bea
&& n'_e (\gamma_e) = \nonumber\\ 
&& \quad N_e \biggl (\frac {\gamma_e}{\gamma_m}\biggr)^{-(p+1)} 
\text {min}[1,\exp(-(\gamma_e - \gamma_i) / \gamma_{cut})] 
\label{nedistpowcut}\\
&& n'_a = \frac{N_e \gamma_m}{p}\biggl [1 - \biggl (\frac{\gamma_i}
{\gamma_m}\biggr )^{-p} + \nonumber\\ 
&& \quad p \biggl (\frac{\gamma_{cut}}{\gamma_m}
\biggr )^{-p} \exp (\frac{\gamma_i}{\gamma_{cut}}) 
\Gamma (-p, \frac{\gamma_i}{\gamma_{cut}})\biggr ] \label{gammacondcut} \\
&& \frac {\gamma'^2 m_p n'_0 \epsilon_e}{m_e} = \frac{N_e \gamma_m^2}{(p - 1)}
\biggl [1 - \biggl (\frac{\gamma_i}{\gamma_m}\biggr )^{-(p-1)} + 
\nonumber \\
&& \quad (p-1)\biggl (\frac{\gamma_{cut}}{\gamma_m}\biggr )^{-(p-1)} 
 \exp (\frac{\gamma_i}{\gamma_{cut}}) \Gamma (-(p-1), 
\frac{\gamma_i}{\gamma_{cut}})\biggr ] \nonumber\\ 
\label{enercondcut}
\eea
where $\Gamma (\alpha,x)$ is the incomplete Gamma function. As the number of 
parameters in this distribution is larger than the number of conservation 
conditions, in contrast to the power-law distribution, it 
is not possible to find an expression for $N_e$ and $\gamma_m$ with respect 
to the total density and the fraction of electric and magnetic energies 
transferred to leptons. A more simplified version of this model is a 
power-law with an exponential cutoff:
\bea
n'_e (\gamma_e) &=& N_e \biggl (\frac {\gamma_e}{\gamma_m}\biggr)^{-(p+1)} 
\exp (-\frac{\gamma_e}{\gamma_{cut}}) \nonumber\\ 
&& \mbox{for } \gamma_e \geqslant \gamma_m \quad , \quad \gamma_{cut} \gg 
\gamma_m \label{nedistpowc} 
\eea
Note that due to the exponential cutoff the restriction to $p \geqslant 2$ 
does not apply and the index of the power-law term can be negative. Using 
conservation conditions (\ref{enum}) and (\ref{eener}) we can find relations 
between parameters of this distribution:
\bea
\gamma_{cut} & = & \frac {\epsilon_e \gamma'^2 m_p n'_0}{|p| m_e n'_a} 
\label{gammacut} \\
\gamma_m N_e & = & n'_a (\frac{\gamma_{cut}}{\gamma_m})^p |\Gamma (-p,
\frac{\gamma_{cut}}{\gamma_m})|^{-1} \label{gammamne}
\eea
As we have only two constraints, it is not possible to find expressions for 
3 constants $N_e$, $\gamma_{cut}$, and $\gamma_m$, and one of them will stay 
as free parameter. In Sec.\ref{sec:synchflux} we show that this type of 
electron distribution is necessary to explain the hard spectrum of the short 
hard and some of the long bursts.

In the introduction we mentioned that the induced transverse magnetic field 
is produced by Weibel instability in the active region. The magnetic energy 
density is parametrized by assuming that it is proportional to the energy 
density of in-falling particles to the shock front/active region:
\be
\frac {B'^2}{8\pi} = \epsilon_B c^2 \gamma' m_p n'_0 \label {magener}
\ee
It is expected that both $\epsilon_e$ and $\epsilon_B$ evolve with 
time/radius. If the central engine is magnetized the external magnetic energy 
can be very important and an external field should be added to the right 
hand side of (\ref{magener}). Here for simplicity we neglect such cases.

Considering the simplest case of a power-law distribution for electrons and 
also assuming that only in-falling electrons are accelerated 
i.e. $n'_a = n'_0 \gamma'$ ($n'_0$ is the minimum of $n'_a$ for a radiative 
shock), the synchrotron term in (\ref{enercons}) and (\ref{momcons}) becomes:
\be
\frac{dE'_{sy}}{4\pi m c^2dr'} = \frac{4 \alpha \sigma_T m_p^2 {n'}_0^2 
\epsilon_e^2 \epsilon_B \gamma'^6 r'^2 \Delta r'}{3 m_e^2 \beta'} , 
\quad \alpha \equiv \frac {(p-1)^2}{p(p-2)} \label{synchr}
\ee
For the reasons explained in the introduction and in Sec. \ref {sec:qualdesc} 
we believe that in a realistic model of relativistic shocks one should 
consider the time evolution of electric and magnetic fields. Here we assume a 
simple power-law evolution with a constant index:
\be
\epsilon_e = \epsilon_e (r'_0) \biggl (\frac {r'}{r'_0} \biggr )^{\alpha_e} 
\quad , \quad \epsilon_B = \epsilon_B (r'_0) \biggl (\frac {r'}{r'_0} 
\biggr )^{\alpha_B} \label {epseb}
\ee
Using (\ref{synchr}), the expression for the column density 
(\ref{columndenssol}), and the momentum conservation equation 
(\ref{momcons}), we obtain the following equation for the evolution of the 
relative Lorentz factor: 
\bea
&&\hspace{-15mm}\frac {d}{dr'}\biggl [\frac {N_0 {r'}_0^3 (\biggl (\frac {r'}
{{r'}_0}\biggr )^{3-\kappa} - 1) + (3-\kappa) {r'}_0^2 n(r'_0)\Delta r' (r'_0)
\gamma'_0 (1-\beta'_0) \beta'}{(3-\kappa) (1-\beta')} \biggr ] = 
\nonumber \\
&& -\frac{4\alpha m_p^2 \sigma_T N_0^2 \epsilon_e^2 (r'_0) \epsilon_B (r'_0) 
\gamma'^6 r'^2 \Delta r' \biggl (\frac{r'}{r'_0}\biggr )^{-\eta}}{3 m_e^2 
\beta'} \label {gammaevol}
\eea
The parameter $\eta \equiv 2\alpha_e + \alpha_B + 2\kappa$ is the evolution 
index of the density and fields. Although this differential equation is of 
order one, it is highly non-linear. To solve it we proceed a perturbative 
method based on iteration. Moreover, it depends explicitly on $\Delta r'$ 
and as we discussed in Sec.\ref{sec:shockevol} to be able to find an explicit 
solution for $\gamma' (r')$, we have to model its evolution. We consider two 
models:

\subsubsection {Dynamically driven active region} \label{sec:dynamicreg}
Assuming that the shock strength and consequently $\Delta r$ depends mainly 
on the density difference, and that the densities of the shells in their rest 
frame are roughly the same, we expect smaller $\Delta r'$ for larger 
$\gamma'$. On the other hand when the relative Lorentz factor is small and 
the shock is soft, $\Delta r'$ should be proportional to $\beta'$ and 
$\Delta r' \rightarrow 0$ when $\beta' \rightarrow 0$. The simplest 
parametrization of $\Delta r'$ with this properties is:
\be
\Delta r' = \Delta r'_0 \biggl (\frac {\gamma'_0 \beta'}{\beta'_0 \gamma'} 
\biggr )^{\tau}\Theta (r'-r'_0) \label {drdyn}
\ee
where $\Delta r'_0$ is a thickness scale. A $\Theta$-function is added to 
(\ref {drdyn}) to explicitly indicate that the expression is valid only 
for $r' \geqslant r'_0$. Note that in this model the initial thickness is not 
null and therefore it is assumed that it was formed in a negligible time or 
the value of $\Delta r'_0$ is the final value from a previous regime that 
makes the initial active region before (\ref{drdyn}) can be applied. This 
model should be suitable for the prompt/internal 
shocks in which two high density narrow shells pass through each other 
and one expect that roughly instantly a narrow and dens active region forms 
around the shock discontinuity (See also next section for other cases). 
As $\beta'/\gamma' < 1$ is expected to be a 
decreasing function of $r'$, for $\tau > 0$ the width of the active region 
decays and for $\tau < 0$ it grows. But it is not always the case, see 
simulations in Paper II.

\subsubsection {Quasi-steady active region} \label{sec:quasistatreg}
At the beginning of a strong shock, presumably an internal shock or when the 
slow shell is extended, roughly homogeneous, and has a low density, we expect 
that after a transient time in which the active region grows, its 
thickness arrives to an stable state determined by the relative Lorentz 
factor, density, synchrotron emission, and expansion of the shells. This 
stability should persist until the loss of kinetic energy due to 
radiation and mass accumulation becomes important, or the fast shell passes 
through the slower one (this does not happen for a radiative shock). 
In this case we parametrize the time evolution of $\Delta r'$ as:
\be
\Delta r' = \Delta r_{\infty} \bigg [1-\biggl (\frac{r'}{r'_0} \biggr )^
{-\delta}\biggr ] \Theta (r-r'_0) \label {drquasi}
\ee
where $\Delta r_{\infty}$ is the final width when the equilibrium is achieved. 
For the decay of the active region at the end of this regime one can use the 
dynamical model. Another possibility is to consider:
\be
\Delta r' = \Delta r_{\infty} \biggl (\frac{r'}{r'_0} \biggr )^{-\delta} 
\Theta (r-r'_0) \label {drquasiend}
\ee
In Appendix \ref{app:a} we argue that with small modifications the 
calculation of dynamical evolution can be used for this model too.

This model is 
specially suitable for studying the external shocks of the ejecta from the 
central source with diffuse material or wind surrounding it, and/or the ISM. 
One expects that in these cases the density of the relativistic ejecta - the 
fast shell - be much higher than the wind or ISM and its extension much 
smaller. In Paper II we show that such a model along with a late emission from 
internal shocks can explain the shallow regime observed in the X-ray light 
curve of the majority of GRBs detected by \swift.

For studying the evolution of the active region, apriori we should also 
take into account the total size of the shells and the passage or coalescence 
time. However, observations show that synchrotron emission continues for a 
significant time after the end of the shock - when shells passed through 
each other or coalesced. For a far observer what is matter is the emission 
rather than physical encounter between shells. All these stages can be 
modelled by one of the models explained here or similar models for the 
evolution of $\Delta r'$. In this case the difference between various stages 
of the collision is reflected in the different value of parameters.

\subsection {Evolution of relative Lorentz factor} \label{sec:lf}
To solve equation (\ref{gammaevol}) we use a perturbative/iterative 
method based on the assumption that the dimensionless coupling in the r.h.s of 
this equation is smaller than one. By dividing both sides of (\ref{gammaevol}) 
with $n_0 {r'}_0^2$ one can extract the coupling ${\mathcal A}$:
\bea
&& \hspace{-5mm}\frac {d}{d\biggl(\frac{r'}{r'_0} \biggr )}\biggl [\frac {(\frac {r'}
{{r'}_0})^{3-\kappa} - 1 + \frac {(3-\kappa) n'(r'_0)\Delta r' 
(r'_0)}{N'_0 r'_0}~\gamma'_0 (1-\beta'_0)) \beta'}{(3-\kappa) (1-\beta')} 
\biggr ] = \nonumber \\
&& \quad -\frac {{\mathcal A} \gamma'^6 \Delta r'}{\beta' \Delta r' 
(r'_0)} \biggl (\frac{r'}{r'_0}\biggr )^{2 -\eta} \label {gammaevolnodim} \\
&& {\mathcal A} \equiv \frac{4\alpha m_p^2 \sigma_T N'_0 \Delta r' (r'_0)
\epsilon_e^2 (r'_0) \epsilon_B (r'_0)}{3 m_e^2} \label {synchcoupling}
\eea
It is straightforward to see that if the initial column density of the slow 
shell/ISM $n'_0 \Delta r' (r'_0) \lesssim 10^{22}$ cm$^{-2}$, for any value of 
$\epsilon_e^2 (r'_0) < 1$ and $\epsilon_B (r'_0) < 1$, the coupling 
${\mathcal A} < 1$. This upper limit on the shell column density is in the 
upper range of the observed total $N_H$ column density of GRBs. However, the 
real $N_H$ can be much higher than what is measured from the absorption of 
the soft X-ray at least $\gtrsim 100$ sec after the trigger, because it is in
conflict with $N_H$ estimated from Lyman-$\alpha$ absorption~\citep{agreion}. 
The difference can be due to the ionization of the neutral hydrogen by UV 
emission from the prompt emission. Nonetheless simulations of the formation 
of the electric and magnetic fields in the shocks show that the fraction of 
the kinetic energy transferred to the fields, specially to the magnetic 
field, is much less than one~\citep{shockmag}. Therefore even with larger 
column densities, the value of ${\mathcal A}$ should be less than one and the 
validity of the perturbative method is justified.

The zero-order approximation corresponds to ${\mathcal A} \rightarrow 0$. 
In this case eq. (\ref{gammaevolnodim}) is a pure differential and its solution 
is trivial:
\be
\beta'_{(0)} (r') = 
\begin{cases}
\frac {(3-\kappa){\mathcal D}}{(\frac {r'}
{{r'}_0})^{3-\kappa} - 1 + \frac {(3-\kappa){\mathcal D}}{\beta'_0}} & 
\kappa \neq 3 \\
\frac {{\mathcal D}}{\ln \frac {r'}{{r'}_0} + \frac {{\mathcal D}}{\beta_0}} 
& \kappa = 3 
\end{cases} \label {betasolzero}
\ee
\be
{\mathcal D} \equiv \frac {n' (r'_0) \Delta r' (r'_0) \beta'_0 \gamma'_0}
{N'_0 r'_0}  \label{dconst}
\ee
where $\beta'_{(0)}$ indicates the zero-order approximation for $\beta'(r')$. 
In the rest of this work we only concentrate on $\kappa \neq 3$, but 
calculations can be easily extended to this exceptional case.

The physical interpretation of (\ref{betasolzero}) is quite evident. 
$\beta' (r')$ changes inversely proportional to the total mass of the shell 
including the accumulated mass of the swept material. This zero-order solution
does not take into account the energy necessary to accelerate particles of 
the slow shell. Thus, its use without radiation corrections will lead to a 
violation of energy conservation. Parameter ${\mathcal D}$ presents the 
strength of the shock; smaller ${\mathcal D}$, faster the constant term in 
the denominator becomes negligible with respect to radial growth and 
$\beta'_{(0)} (r')$ approaches a cubic decline (for $\kappa = 0$) due to the 
adiabatic expansion. The origin of term $(3 - \kappa)$ is partially 
geometrical and partially related to the density variation with $r$ in the 
slow shell. It is the effective mass accumulation index of the shock.

As (\ref{betasolzero}) is the dominant component of the dynamics and is 
used through out this work, it is useful to have its asymptotic behaviour for 
$(\frac {r'}{{r'}_0}) \gtrsim 1$ and $(\frac {r'}{{r'}_0}) \gg 1$:
\be
\beta'_{(0)} (r') \approx 
\begin{cases}
\frac {{\mathcal D}}{\varepsilon + \frac {{\mathcal D}}{\beta'_0}} \approx 
\beta'_0 (1 - \frac {\beta'_0 \varepsilon}{{\mathcal D}}) & 
\frac {r'}{{r'}_0} - 1\equiv \varepsilon \gtrsim 0 \\
(3-\kappa){\mathcal D}(\frac {{r'}_0}{r'})^{3-\kappa} & \mbox {For } 
(\frac {r'}{{r'}_0})^{3-\kappa} \gg \frac {(3-\kappa){\mathcal D}}
{\beta'_0} \\
\beta'_0 (1 - \frac {\beta'_0}{(3 - \kappa){\mathcal D}}) 
(\frac {r'}{{r'}_0})^{3-\kappa} & \frac {(3-\kappa){\mathcal D}}
{\beta'_0} > (\frac {r'}{{r'}_0})^{3-\kappa} \gg 1
\end{cases} \label {betasolzeroapp}
\ee
We use the zero-order solution in the r.h.s. of (\ref{gammaevolnodim}) to 
obtain the first-order correction of the solution. The $n$-order approximation 
corresponds to using the $(n-1)$-order approximation in the r.h.s. of 
(\ref{gammaevolnodim}) and solving the equation:
\bea
&& \frac {(\frac {r'}{{r'}_0})^{3-\kappa} - 1 + \frac {(3-\kappa) 
n'(r'_0)\Delta r' (r'_0)}{N'_0 r'_0}~\gamma'_0 (1-\beta'_0)) \beta'_{(n)}}
{(3-\kappa) (1-\beta'_{(n)})} = \nonumber \\
&&\quad {\mathcal D}_{(n)} -\frac {{\mathcal A}}{\Delta r' (r'_0)} \int_1^{\frac{r'}
{r'_0}}\frac {{\gamma'}_{(n-1)}^6 \Delta r'}{\beta'_{(n-1)}} x^{2 -\eta}dx 
\label{betansol}
\eea
where ${\mathcal D}_{(n)}$ is an integration constant. For any order of 
correction (\ref{betansol}) must satisfy the initial condition i.e. 
$\beta'_{(n)}(r'_0) = \beta'_0$ and from this constraint one can determine 
${\mathcal D}_{(n)}$. It is easy to see that ${\mathcal D}_{(n)} = 
{\mathcal D}$ for all orders of perturbation. Calling the integral term 
${\mathcal M}_{(n-1)}(r')$ for $(n-1)$-order solution, we find the following 
recursive expression for $\beta'_{(n)}$:
\be
\beta'_{(n)} = \frac {{\mathcal D} - \frac { {\mathcal A}{\mathcal M}_{(n-1)}
(r')}{\Delta r' (r'_0)}}{\frac {1}{3-\kappa}((\frac {r'}{{r'}_0})^
{3-\kappa} - 1) + \frac {{\mathcal D}}{\beta'_0} - 
\frac {{\mathcal A} {\mathcal M_{(n-1)}}(r')}{\Delta r' (r'_0)}} \label {betan}
\ee
At this point we have to consider a model for $\Delta r'$. For the 
dynamical model ${\mathcal M}_{(0)}$ becomes:
\bea
{\mathcal M}_{(0)}(r') & = & \Delta r'_0 \biggl (\frac{\gamma'_0}{\beta'_0} 
\biggr )^{\tau} \int_1^{\frac{r'}{r'_0}}\frac {{\beta'_{(0)}}^{\tau - 1}}
{(1 - {\beta'_{(0)}}^2)^{3-\frac {\tau}{2}}}~x^{2 -\eta} dx \nonumber \\
& = & - \frac {\Delta r'_0}{3-\kappa} \biggl (\frac{\gamma'_0}{\beta'_0} 
\biggr )^{\tau} ((3-\kappa) {\mathcal D})^{1-\frac{\eta}{3-\kappa}} 
\int_{\beta'_0}^{\beta'_{(0)} (r')} dy \nonumber \\
& & \frac{y^{\tau - 3 + \frac{\eta}{3-\kappa}}}{(1 - y^2)^{3 - 
\frac{\tau}{2}}} (1 + (\frac {1}{(3-\kappa) {\mathcal D}} - \frac {1}
{\beta'_0}) y)^{-\frac{\eta}{3-\kappa}} \nonumber \\
\label {integdyn}
\eea
The second form of the integral is obtained using $\beta'_{(0)}$ solution. 
Unfortunately this integral can not be determined analytically. Nonetheless, 
by expanding one of the two terms in the integrand it is possible to find an 
approximation which is useful for getting an insight into 
the behaviour of $\beta'_{(1)}$, its dependence on various parameters, and the 
importance of radiation correction. They are summarized in Appendix 
\ref{app:a}.

For quasi steady model ${\mathcal M}_{(0)}(r')$ is:
\bea
&& {\mathcal M}_{(0)}(r') = \Delta r'_{\infty} \int_1^{\frac{r'}{r'_0}}
\frac {x^{2 -\eta} (1-x^{-\delta})}{(1 - {\beta'_{(0)}}^2)^3} dx = \nonumber \\
&& \quad -\frac {\Delta r'_{\infty}}{3-\kappa} ((3-\kappa) {\mathcal D})^
{1-\frac{\eta}{3-\kappa}} \int_{\beta'_0}^{\beta'_{(0)} (r')} dy \nonumber \\
&& \quad \frac{y^{\frac{\eta}{3-\kappa} - 3}}{(1 - y^2)^{3}} (1 + (\frac {1}
{(3-\kappa) {\mathcal D}} - \frac {1}{\beta'_0}) y)^{-\frac{\eta}{3-\kappa}} 
\nonumber \\
&& \biggl [ 1 - ((3-\kappa) {\mathcal D})^{-\frac{\delta}{3-\kappa}} 
(1 + (\frac {1}{(3-\kappa) {\mathcal D}} - \frac {1}{\beta'_0}) y)^
{-\frac{\delta}{3-\kappa}} \biggr ]\nonumber \\
\label {integsteady}
\eea
The similarity of the second term in (\ref{integsteady}) to the integrand 
of (\ref{betan}) shows that at first order of radiative correction the two 
terms in (\ref{drquasi}) act independently. The constant term leads to the 
first term in (\ref{integsteady}) which is 
equivalent to (\ref{betan}) with $\tau \rightarrow 0$, i.e. an active 
region with constant thickness. The term proportional to $(r'_0/r')^{\delta}$ 
in (\ref{drquasi}) is responsible for the second term in (\ref{integsteady}). 
Up to a constant it is also the expression for a constant thickness model 
with $\eta \rightarrow \eta + \delta$. This means that the effect of power-law 
term in (\ref{drquasi}) is very similar to power-law dependence of fields and 
shell density on $r'$. This behaviour justifies the name 
{\it quasi-steady} we have given to this model. Analytical approximations 
of these integrals can be found in Appendix \ref{app:a}. 

In the last paragraph we considered power-law dependence on $r'$ for the 
electric and magnetic 
fields. The case of an exponential rise or fall of the fields is important at 
the beginning and at the end of gamma-ray spikes or early X-ray flares. For 
this case, the only modification in (\ref{integdyn}) and (\ref{integsteady}) 
is the replacement of $x^{-\eta}$ with $\exp (-\eta x)$ where here $\eta$ is a 
dimensionless coefficient determining the speed of exponential variation. It 
is negative for rising fields and positive when fields are declining, similar 
to the power-law case. In the same way if the electron distribution 
$n'_e (\gamma_e)$ includes an exponential cutoff~\citep{ecutoff}, an 
exponential term similar to the term for fields appears in the expression for 
${\mathcal M}_{(0)}(r')$. Therefore in a general case, the integrand in 
(\ref{integdyn}) and (\ref {integsteady}) includes an exponential term 
which makes it even more complex. Nonetheless, the expansion of the 
exponential permits to obtain the analytical approximation given in the 
Appendix \ref{app:a}. 

Finally by using ${\mathcal M}_{(0)}(r')$ in (\ref {betan}) and 
(\ref{columndenssol}), we can determine $\beta'_{(1)}$ the first-order 
radiation corrected evolution of $\beta'$ and column density $n'(r')\Delta r'$ 
with radius/time. The complexity of expressions for ${\mathcal M}_{(0)}(r')$ 
and consequently for $\beta'_{(1)}$ does not permit to investigate the effect 
of various quantities from the exact calculations and we leave this for 
Paper II where we numerically evaluate the behaviour of shocks kinematic 
and radiation. Here we just consider the simplest cases when in 
(\ref {betasolzeroapp}), $r'/{r'}_0 \gtrsim 1$ or $r'/{r'}_0 \gg 
(3 - \kappa) {\mathcal D} / \beta'_0$. Using (\ref{dynmzerofirst}) and 
(\ref{dynmzerofirsteta}) respectively for large and small $\eta$, we find 
following expressions for ${\mathcal M}_{(0)}(r')$ when $\varepsilon \ll 1$:
\bea
{\mathcal M}_{(0)}(r') & \approx & \Delta {r'}_0 \eta{\mathcal B}
(1-\frac{1}{{\mathcal C}})(1 + (3-\frac{\tau}{2}) \beta_0^2)
\varepsilon \nonumber \\ 
&&\mbox{for }|\eta| \gg 0 \label {largeetaeps}
\eea
\bea
{\mathcal M}_{(0)}(r') & \approx & \Delta {r'}_0 {\mathcal B}
 \biggl [(3-\kappa)(\tau - 2 + \frac{\eta}{3-\kappa}) + \nonumber \\ 
&& {\beta'_0}^2 \biggl (\frac {(3-\kappa)(3-\frac{\tau}{2})(\tau - 1 + 
\frac{\eta}{3-\kappa})}{(\tau + \frac{\eta}{3-\kappa})} - \nonumber \\ 
&& \frac{\eta (3-\frac{\tau}{2})(1 - \frac{1}{{\mathcal C}})}{\tau + 1 + 
\frac{\eta}{3-\kappa}} \biggr ) \biggr ] \varepsilon \quad \quad 
\mbox{for }|\eta| \gtrsim 0 \label {smalletaeps}\\
{\mathcal B} & \equiv & \frac{{\gamma'_0}^\tau}{(3-\kappa)\beta'_0}~ 
{\mathcal C}^{-\frac{\eta}{3-\kappa}} \quad , \quad 
{\mathcal C} \equiv \frac{(3-\kappa){\mathcal D}}{\beta'_0} \label{bdef}
\eea
Therefore ${\mathcal M}_{(0)}(r') \propto \Delta {r'}_0 {\mathcal B} 
\varepsilon$ for $\varepsilon \ll 1$. The constant coefficient is expected 
to be of order 1. By applying these results to (\ref{betan}) we find:
\be
\beta'_{(1)} \approx \frac{{\mathcal D} - {\mathcal A}{\mathcal B}'
\varepsilon}{\frac {{\mathcal D}}{\beta'_0} + (1 - {\mathcal A}{\mathcal B}') 
\varepsilon} \label {betaone}
\ee
where ${\mathcal B}'$ is ${\mathcal B}$ multiplied by the corresponding 
constant coefficients in (\ref{largeetaeps}) or (\ref{smalletaeps}) depending 
on the value of $\eta$. Comparing this result with the corresponding 
$\beta'_{(0)}$ we conclude that the strength of the radiation correction 
of $\beta'$ and its effect on the kinematic of the ejecta/jet depends on 
${\mathcal S} \equiv {\mathcal A}{\mathcal B}'$. As ${\mathcal B}'$ is 
proportional to 
${\mathcal D}^{-\frac{\eta}{3-\kappa}}$, for a positive $\eta$ and same 
${\mathcal A}$, larger ${\mathcal D}$ (stronger shock), smaller 
${\mathcal S}$. In this case the kinetic energy of the shock is much larger 
than radiation and therefore the synchrotron emission does not significantly 
modify the evolution of the shock. Note also that although ${\mathcal S}$ is 
linearly proportional to the synchrotron total coupling ${\mathcal A}$, it 
depends non-linearly on ${\mathcal D}$ through a power which depends on the 
time/radius variation of the electric and magnetic fields as well as the 
density of the shells. The quantity ${\mathcal S}$ in this model depends 
also on $\gamma'_0$: larger $\gamma'_0$, larger the influence of radiation.
This simply means that the effective thickness of the active region decreases 
faster when the effect of radiation is stronger. This behaviour is consistent 
with the phenomenology of the model described here.

Similar expressions can be found for the other extreme case i.e. when 
$r'/{\mathcal C}r'_0 \gg 1$:
\bea
&& {\mathcal M}_{(0)}(r') \approx \frac {\Delta {r'}_0 {\gamma'_0}^\tau \eta 
{\mathcal C}^{1 - \frac{\eta}{3-\kappa}} (1- \frac{1}{{\mathcal C}})}
{(3-\kappa)^2} \biggl \{[\frac {1}{\tau - 1 + \frac{\eta}{3-\kappa}} + 
\nonumber \\
&&\quad \frac {(3-\frac{\tau}{2}) {\beta'_0}^2}{\tau + 1 + \frac{\eta}{3-
\kappa}}] - \biggl(\frac{{\mathcal C} {r'_0}^{3-\kappa}}{r'^{3-\kappa}} 
\biggr )^{\tau - 1 + \frac{\eta}{3-\kappa}}[\frac {1}{\tau - 1 + \frac{\eta}
{3-\kappa}} + \nonumber \\
&&\quad \frac {(3-\frac{\tau}{2}) {\beta'_0}^2}{\tau + 1 + 
\frac{\eta}{3-\kappa}} \biggl(\frac{{\mathcal C} {r'_0}^{3-\kappa}}{r'^
{3-\kappa}} \biggr )^2] \biggr \} \quad \quad \mbox{for }|\eta| \gg 0 
\label {largeetar}
\eea
\bea
&& {\mathcal M}_{(0)}(r') \approx \frac {\Delta {r'}_0 {\gamma'_0}^\tau 
{\mathcal C}^{1 - \frac{\eta}{3-\kappa}} (1- \frac{1}{{\mathcal C}})}
{(3-\kappa) \beta'_0}\biggl \{[1 + \frac {(3-\frac{\tau}{2}) {\beta'_0}^2}
{\tau + \frac{\eta}{3-\kappa}} + \nonumber \\
&& \quad \frac{\eta (1 - \frac{1}{{\mathcal C}})}{3-\kappa}(1+\frac {(3-
\frac{\tau}{2}) {\beta'_0}^2}{\tau + 1 + \frac{\eta}{3-\kappa}})] - 
\biggl(\frac{{\mathcal C} {r'_0}^{3-\kappa}}{r'^{3-\kappa}} \biggr )^
{\tau - 2 + \frac{\eta}{3-\kappa}} \nonumber \\
&& \quad [1 + \frac {(3-\frac{\tau}{2}) {\beta'_0}^2 
\biggl(\frac{{\mathcal C} {r'_0}^{3-\kappa}}{r'^{3-\kappa}} \biggr )^2 }
{\tau + \frac{\eta}{3-\kappa}} + \frac{\eta \frac{{\mathcal C} {r'_0}^
{3-\kappa}}{r'^{3-\kappa}}(1 - \frac{1}{{\mathcal C}})}{3-\kappa}(1+ 
\nonumber \\
&& \quad \frac {(3-\frac{\tau}{2}) {\beta'_0}^2 \biggl(\frac{{\mathcal C} 
{r'_0}^{3-\kappa}}{r'^{3-\kappa}} \biggr )^2}{\tau + 1 + \frac{\eta}
{3-\kappa}})] \biggr \} \nonumber \\
& &\quad \quad \mbox{for }|\eta| \gtrsim 0 
\label {smalletar}
\eea
When $r'/{\mathcal C}r'_0 \rightarrow 0$, ${\mathcal M}_{(0)}(r') \rightarrow 
const.$, if $\tau - 2 + \frac{\eta}{3-\kappa} > 0$ for $|\eta| \gtrsim 0$ or 
$\tau - 1 + \frac{\eta}{3-\kappa} > 0$ for $|\eta| \gg 1$. This can be 
interpreted as a saturation state for the synchrotron radiation which has been 
observed specially in bright bursts where peaks of the prompt gamma-ray 
emission are roughly square-like. Examples are GRB 060105~\citep{grb060105}, 
GRB 061007~\citep{grb061007,grb061007-1}, GRB 060813A~\citep{grb060813a}, and 
GRB 070427~\citep{grb070427}. Even the {\it Super Burst} 
GRB 080319B~\citep{grb080319b,grb080319b1} seems to consist of 3 overlapping 
square shape peaks. See also simulations in Paper II.

The conditions mentioned above can be considered as consistency conditions 
because if they are not satisfied ${\mathcal M}_{(0)}(r') \rightarrow \infty$ 
which is not physically acceptable. Therefore these conditions constrain 
parameters of the model - $\eta$, $\tau$, and $\kappa$. For instance, for a 
slow shell with a roughly constant density $\kappa \sim 0$. Therefore $\eta$ 
depends only on the behaviour of electric and magnetic fields. If the  
variation index of the these fields are small and positive, the value of 
$\tau$ can be small, i.e. the radiation will not vary very quickly. By 
contrast, 
a negative index - increasing fields - can not last for long time and 
imposes a large value for $\tau$. This simple argument shows that there is 
an intrinsic relation between these parameters. However, only a detailed 
modelling of the microphysics of the shock will be able to determine possible 
relations and their physical origin. Nonetheless the discussion above shows 
that the simple model studied here is consistent and includes some of the 
important properties of the phenomena involved in the production of GRBs.

Using (\ref{betan}) one can see that a constant ${\mathcal M}_{(0)}(r')$ is 
equivalent to redefinition of ${\mathcal D}$. Therefore when the radiation 
term arrives to its maximum, $\beta'$ evolution becomes like a non-radiating 
ejecta.

In the case of a quasi-steady active region, the behaviour of 
${\mathcal M}_{(0)}(r')$ is essentially similar because of similarity between 
two models explained above. However, the time scales and indexes are 
different. For instance, when $\varepsilon \ll 1$ the value of 
${\mathcal M}_{(0)}(r')$ is proportional to $\varepsilon$ with a coefficient 
equal to (\ref{largeetaeps}) or (\ref{smalletaeps}) and $\tau = 0$, minus the 
same term with $\eta \rightarrow \eta + \delta$. In this case, the initial 
$\gamma'_0$ does not have an explicit contribution and the slope of 
${\mathcal M}_{(0)}(r')$ is smaller than dynamical model. For 
$r'/{\mathcal C}r'_0 \gg 1$ if other parameters are the same as dynamical 
model, the absence of $\gamma'_0$ and smaller power of ${\mathcal C}r'_0/r'$ 
means that ${\mathcal M}_{(0)}(r')$ approaches its maximum value slower. 
There is also a slower change when one of the two contributor terms becomes 
too small and negligible. As expected, all these properties make this model 
more suitable for modelling the afterglow.

Up to now we have only discussed the solutions of dynamical equation 
(\ref{gammaevol}) corresponding to the rise of the synchrotron emission. The 
falling edge of the emission i.e. when ${\mathcal M}_{(0)}(r') \rightarrow 0$ 
can be obtained simply be time/radius reversal of rising solutions, 
see eq. (\ref{drquasiend}). For 
instance in (\ref{largeetaeps}) and (\ref{smalletaeps}) if $r' < r'_0$, 
$\varepsilon < 0$. By moving the initial condition from $r'_0$ to $r'$, we 
obtain a positive but decreasing value for ${\mathcal M}_{(0)} (r')$ which 
becomes zero at $r' = r'_0$. These approximations however do not permit 
to determine when the radiation begins to decrease. For this we need a 
detailed study of the evolution of fields and other shock properties.

In summary, the evolution of $\beta'$ determines the kinematic of the 
burst and is important for the estimation of all observables such as what we 
will discuss in the next sections - synchrotron flux, hardness ratios, etc. 
$\beta'$ is also important for 
determining the evolution of 
other parameters that are not directly observable and a model must be used for 
their extraction from data. A good example is the time variation of 
$\omega'_m$ the minimum characteristic frequency of the synchrotron emission. 
It determines the behaviour of the spectrum and light curves (see expression 
(\ref{omegamdef}) below for its definition). Assuming the simplest case of 
$n'_a = n'_0$ in (\ref{gammam}), from the definition of $\omega'_m$ one can 
see that $\omega'_m \propto \gamma'^2 \epsilon_B$. The proportionality 
coefficient is time/radius independent. 

\subsection {Synchrotron flux and spectrum} \label{sec:synchflux}
In this section we first remind the synchrotron emission for the purpose of 
completeness and then we use the results for determination of lags between 
the light curves of different energy bands.

The ejecta from a central engine that produces the gamma-ray burst is most 
probably collimated and jet-like, otherwise the observed energy is not 
explainable. On the other hand for the far observers, even a spherical 
relativistic emission looks collimated to an angle $\theta < 1/\Gamma$ along 
the line of sight where $\Gamma$ is the Lorentz factor of the emitting 
matter in the observer rest frame~\citep{relcollim}. Therefore in any case 
we need to consider the angular dependence of the synchrotron emission. 
Moreover, we need to consider the delay as well as angular dependence of the 
Doppler shift of the emission. Simulations show that even with an angular 
independent emission, these effects can apriori explain the lag between 
different bands observed in both BATSE and \swift 
bursts~\citep{dopplerlag,dopplerlag0,dopplerlag1}. However, in these 
simulations the spectrum and the time profile of emission have been put by 
hand. 

There are a number of evidence against a high latitude/Doppler effect origin 
of the observed lags. First of all the total effect of high latitude emission 
decreases when Lorentz factor is very high. It is expected that in GRBs 
$\Gamma \gtrsim 100$, and therefore this effect should be very small. In 
addition, even in the early X-ray emission where it was expected that 
this effect dominates, it has not been observed. On the other hand, for a 
given category of bursts, short or long, it does not seem that there is any 
relation between lags and spectrum as expected from a Doppler effect. 
Therefore we conclude that the contribution of high latitude emission and 
Doppler effect is sub-dominant and can not explain the observations. 
On the other hand, we show that even with neglecting the high latitude 
emission, due to the evolution of physical properties a lag between different 
energy bands exists.

In the model presented here the synchrotron emitting matter is confined to 
the active region. Therefore we identify the bulk Lorentz factor $\Gamma$ 
with respect to the observer with the average Lorentz factor of the active 
region with respect to the observer. It can be related to the relative 
Lorentz factor $\gamma' (r')$ obtained in Sec. \ref{sec:shockevol} and the 
final Lorentz factor when two shells are coalesced:
\be
\Gamma (r) = \Gamma_f \gamma' (r)(1 + \beta_f \beta' (r)) \label {gammabulk}
\ee
where $\Gamma_f$ is the Lorentz factor of coalesced shells with respect to 
observer. The initial value of $\gamma' (r)$ is the relative Lorentz factor 
of the colliding shells. In the case of an external shocks on a low velocity 
surrounding material or the ISM $\Gamma_f \approx 1$ and $\Gamma (r) \approx 
\gamma'(r)$. Note that here we have written $\gamma$ with respect to observer 
coordinate $r$ because on what concerns the synchrotron emission, only the 
observations of far observers matter.

The energy (intensity) angular spectrum of synchrotron 
emission~\citep{electrody} by one electron or positron in a frame where it 
is accelerated to a Lorentz factor of $\gamma_e$ is\footnote{The expression 
(\ref {synchintens}) is valid for small $\theta'$ angles. However, the main 
part of the emission is in $\gamma_e \theta' < 1$, and the intensity 
exponentially decreases for $\gamma_e \theta' \gtrsim 1$. Therefore it is a 
good approximation for any angle~\citep{electrody}.}:
\bea
\frac {d^2 I'}{d\omega' d\Omega'} & = & \frac {3e^2 {\omega'}^2 \gamma_e^2}
{4\pi^2 {\omega_c'}^2 c}~(1 + \gamma_e^2 {\theta'}^2)^2 \biggl 
[K_{2/3}^2 (\zeta) + \nonumber \\
&& \frac {{\gamma_e^2 \theta'}^2}{1 + \gamma_e^2 
{\theta'}^2}~K_{1/3}^2 (\zeta) \biggr ] \label {synchintens}\\
\omega_c' & \equiv & \frac {3}{2} \gamma_e^3 \biggl (\frac {c}{\rho'} 
\biggr ) = \frac {3e \gamma_e^2 B'}{2c m_e} \label {synchcharact} \\
\zeta & \equiv & \frac {\omega'}{2 \omega_c'} (1 + \gamma_e^2 {\theta'}^2)^
{\frac {3}{2}} \label {zetadef}
\eea
Quantities with a prime in (\ref {synchintens}) to (\ref {zetadef}) are with 
respect to the frame where Lorentz factor of electrons is $\gamma_e$. Here we 
identify this frame as the rest frame of the active region. $\rho'$ is the 
Larmor radius of electrons. The angle $\theta'$ is the angle between 
acceleration direction and emission. Without loss of generality it 
can be assumed to be in x-z plane. Therefore $d\Omega = \cos \theta d\theta 
d\phi$~\citep{electrody}. When $|\theta| \gtrsim 0$, 
$d\Omega \approx d\theta d\phi$.

To obtain the power spectrum that is the measured quantity, we must divide 
the intensity by the precession period of electrons $2\pi \rho' / c$. We 
expect that accelerated electrons have a range of Lorentz factors, therefor 
we should also integrate over their distribution:
\bea
\frac {d^2 P'}{\omega' d\omega' d\Omega'} &=& \frac {e^2} {4\pi^3 c} 
\int_{\gamma_m}^\infty d\gamma_e~n'_e (\gamma_e) \gamma_e^{-1}
\biggl (\frac {\omega'}{\omega_c'}\biggr )~(1 + 
\gamma_e^2 {\theta'}^2)^2 \nonumber \\
&& \biggl [K_{2/3}^2 (\zeta) + \frac {{\gamma_e^2 
\theta'}^2}{1 + \gamma_e^2 {\theta'}^2}~K_{1/3}^2 (\zeta) \biggr ] 
\label {synchpowp}
\eea
Note that we have divided the angular power spectrum by $\omega'$ to make 
it dimensionless. In the observer's rest frame the spectrum is transferred 
as~\citep{pow,pow1,pow2}:
\bea
\frac {d^2 P}{\omega d\omega d\Omega} & = & \frac {1}{\Gamma^2 (1 - 
{\mathbf \beta} \cos \theta)^2}~\frac {d^2 P'}{\omega' d\omega' d\Omega'} 
\label {synchpow} \\
\omega & = & \frac {\omega'}{\Gamma (1 - {\mathbf \beta} \cos \theta)} 
\label {enerdoppler}
\eea
\be
\cos \theta  = \frac {\cos \theta' + {\mathbf \beta}}{1+{\mathbf \beta}
\cos \theta'} \quad , \quad \cos \theta' = \frac {\cos \theta - 
{\mathbf \beta}}{1-{\mathbf \beta}\cos \theta} \label {costheta}
\ee 
where in equations (\ref{synchpow}) to (\ref{costheta}), ${\mathbf \beta}$ is 
related to $\Gamma$ the Lorentz factor of the active region with respect to 
the observer. To find the total power at a given frequency $P (t, \omega)$ in  
the frame of a far observer, we integrate over the distribution of 
accelerated electrons and the emitting volume but constrain it to the 
emission in the direction of the observer. As we assumed that the active 
region is thin, we neglect the absorption of synchrotron photons inside the 
active region itself. Without 
loss of generality we put the observer at $\Omega = \Omega' = 0$. In this case 
for an electron moving at angle $\Omega'_1$ with respect to observer, only 
photons emitted in the direction of $\Omega' = \Omega'_1$ are detected by the 
observer. Therefore we need to integrate either on $\Omega'_1$ 
(or equivalently $\Omega_1$) or on $\Omega'(\Omega)$. For simplicity of 
notation we use the latter. As the synchrotron angular distribution 
does not depend on $\phi$ (or $\phi'$) we only need to integrate over 
$\theta$ (or equivalently $\theta'$):
\be
\frac {dP}{\omega d\omega} \equiv 2\pi \int_r^{r + \Delta r} dr 
\int_{\theta_{min}}^{\theta_{max}} d\theta \cos \theta \frac {d^2 P (t - 
\Delta t), \theta, \omega)}{\omega d\omega d\Omega} \label {totpow}
\ee
$\theta_{min}$ and $\theta_{max}$ are minimum and maximum visible angle for 
the observer with the constraint $|\theta_{min}| < 1/\Gamma$ and 
$|\theta_{max}| < 1/\Gamma$. We define $\Delta \theta \equiv |\theta_{max} + 
\theta_{min}| / 2$ as the view angle of the observer with respect to the 
ejecta/jet axis. These angles are not directly measurable and therefore the 
simplest assumption is a symmetric ejecta $\theta_{max} = -\theta_{min} = 
1/\Gamma$, i.e. $\Delta \theta = 0$. Note that photons coming from 
$\theta \neq 0$ arrive to the observer with a time delay $\Delta t$ where 
$\Delta t (\theta)$ is~\citep{radproc}:
\be
\Delta t(\theta) = \frac{r(1-\cos \theta)}{c {\mathbf \beta} (r)(1 + 
{\mathbf \beta}^2 \Gamma^2 \sin^2 \theta)^{\frac {1}{2}}} \label {timedelay}
\ee
In (\ref{timedelay}) we have assumed that the initial radius from which the 
shells are ejected from the central engine is much smaller than their 
distance from it when they collide. Therefore, the initial radius is neglected. 
Due to the direct relation between radius and time in this model, 
$t-\Delta t(\theta)$ can be replaced by 
$\Delta r (\theta) = r - c {\mathbf \beta (r)} \Delta t(\theta)$. The quantity 
$\Delta r (\theta)$ should not be confused with the thickness of the active 
region $\Delta r$.

In this model we have considered $\Delta r \ll r$ and the physical 
properties of the active region are close to uniform. Thus, the integral 
over the interval $r$ and $r+\Delta r$ becomes trivial if we consider  
$r$ to be the average distance of the active region. This is similar to the 
way we calculated kinematic quantities in Sec.\ref{sec:shockevol}. With this 
simplification the total spectrum becomes:
\be
\frac {dP}{\omega d\omega} \equiv 2\pi\Delta r \int_{\theta_{min}}^
{\theta_{max}} d\theta \cos \theta \frac {d^2 P (r - \Delta r (\theta), 
\theta, \omega)}{\omega d\omega d\Omega} \label {totpowdeltar}
\ee
If we use (\ref{revol}) to describe $r$ as a function of time, equation 
(\ref {totpowdeltar}) depends only on $t$ and $\omega$.

The width $\Delta r$ can in general depend on the energy. In 
Sec.\ref{sec:qualdesc} we 
described that when electrons lose their energy, they get distance from the 
shock front or in another word are pushed to the upstream. Although in this 
region the magnetic field is expected to be weaker, it can be enough for 
the low energy emissions, UV, optical, IR. Moreover, in a collimated or 
structured jet\footnote{A structured jet is usually considered to have a 
transverse gradient in density~\citep{structedjet,structedjet0}. However it is 
expected that gradually a transverse gradient in Lorentz factor forms too.} 
these less accelerated particles emit mostly in lower energies. In this 
simple model of shock one way of taking into these effects is to consider 
that $\Delta r$ as well 
as $\theta_{min}$ and $\theta_{max}$ depend on the energy. For $\Delta r$ we 
can simply assume that $\Delta r_0$ in dynamical model or $\Delta r_\infty$ 
in the quasi-steady model are energy dependent. Estimation/modeling of energy 
dependence of $\theta_{min}$ and $\theta_{max}$ is more difficult because 
the only way to modify them is through $\Gamma$. Assuming 
$\theta_{jet} > 1/\Gamma$ we must consider that $\Gamma$ is 
$\theta$-dependent. 
This needs a modification of the dynamics and makes the model too complicated.
For this reason here we ignore the energy dependence of opening angle of 
the jet.

The differential term in (\ref{totpowdeltar}) can be replaced by 
(\ref{synchpow}) and (\ref{synchpowp}). We must also take into account the 
proper time delay discussed above. Therefore:
\bea
\frac {dP}{\omega d\omega} & = & \frac{e^2}{2\pi^2} r^2 \Delta r 
\biggl (\frac{\omega'}{\omega'_{cc}} \biggr )\int_{\theta_{min}}^
{\theta_{max}} d\theta \cos \theta \biggl [\Gamma^2 (r - \nonumber \\
&& \frac{r(1-\cos \theta)}{(1+\beta^2\Gamma \sin^2\theta)^{1/2}})~(1 - 
\beta (t) \cos \theta)^2 \biggr ]^{-1} \nonumber \\
& & \int_{\gamma_m}^\infty d\gamma_e \frac {n'_e (\gamma_e)}{\gamma_e^3} 
(1 + \gamma_e^2 {\theta'}^2)^2 \biggl [ K_{2/3}^2 (\zeta) + \nonumber \\
&& \frac {{\gamma_e^2 \theta'}^2}{1 + \gamma_e^2 {\theta'}^2}~K_{1/3}^2 
(\zeta) \biggr ] \label {tpowinter}\\
\omega'_{cc} & \equiv & \frac {\omega'_c}{\gamma_e^2} = \frac {3eB'}{2c m_e}
\label{omegacc}
\eea
As the angle $\theta$ and therefore the delay is small, we use Taylor expansion 
around $r$ to obtain the explicit expression for $\Gamma$ and $\beta$ in the 
integrand of (\ref{tpowinter}):
\bea
&&\Gamma (\frac{r(1-\cos \theta)}{(1+\beta^2\Gamma^2 \sin^2 \theta)^{1/2}}) 
\approx \Gamma (r) (1 - \nonumber \\
&& \quad \quad \quad \quad \frac {r (1-\cos \theta)}{(1 + \beta^2\Gamma^2
\sin^2 \theta)^{1/2}}~\frac{d\Gamma (r)}{\Gamma dr} + \ldots) \label {gammaexp}
\eea
In a similar way we can expand $\beta (r, \theta)$ around $r$:
\bea
&& \beta (\frac{r(1-\cos \theta)}{(1+\beta^2\Gamma^2 \sin^2 \theta)^{1/2}}) 
\approx \beta (r) (1 - \nonumber \\
&& \quad \quad \quad \quad \frac {r (1-\cos \theta)}{(1+\beta^2\Gamma^2 
\sin^2 \theta)^{1/2}}~\frac{d\beta (r)}{\beta dr} + \ldots) \label {betaexp}
\eea
As $|\sin \theta| \sim \theta \leqslant 1/\Gamma$, we can also expand the 
$\theta$ dependent terms in (\ref{gammaexp}) and (\ref{betaexp}). It is 
more convenient to use the relation between $\theta$ and $\theta'$ in 
(\ref{costheta}) and transfer $\theta$ to $\theta'$. We keep 
only terms up to $\theta'^2$ order. With these simplifications the total 
energy spectrum can be written as:
\bea
\frac {dP}{\omega d\omega} & = & \frac{e^2}{2\pi^2} r^2 
\biggl (\frac{\omega'}{\omega'_{cc}} \biggr )\frac{\Delta r}{\Gamma (r)}
\int_{\theta'_{min}}^{\theta'_{max}} d\theta' (\cos \theta' + \beta) \nonumber \\
&& \biggl [1 + {\mathcal G}(r) \theta'^2 + \ldots  
\biggr ] \int_{\gamma_m}^\infty d\gamma_e \nonumber \\
&& \frac {n'_e (\gamma_e)}{\gamma_e^3} 
(1 + \gamma_e^2 {\theta'}^2)^2 \biggl [K_{2/3}^2 (\zeta) + \nonumber \\
&& \frac {{\gamma_e^2 \theta'}^2}{1 + \gamma_e^2 {\theta'}^2}~K_{1/3}^2 
(\zeta) \biggr ] \label {totpowinter} \\ 
{\mathcal G}(r) & \equiv & \frac {{\beta'}^2(r) (1-\beta)}{2(1+\beta)}~
\biggl (\frac{\beta'_0 - \beta'(r)}{\beta'_0 \beta' (r)} + 
\frac{1}{{\mathcal D}}\biggr ) \nonumber \\
&& \biggl (\frac{\beta_f}{1+\beta_f \beta'(r)} - \gamma'^2 \beta'(r) \biggr) 
\label{thetacoef}
\eea
The coefficient ${\mathcal G}(r)$ in (\ref{totpowinter}) presents the lowest 
order correction due to the Doppler effect and delay of high latitude 
emissions. As expected, for $\Gamma \rightarrow \infty$, 
${\mathcal G}(r) \propto 1/\Gamma^2 \rightarrow 0$. Moreover, 
${\mathcal G}(r)$ does not depend on $\omega'$ (or equivalently $\omega$), 
and therefore it does not affect the lag between different energy bands 
itself. 

Apart from the energy dependence of $\Delta r$, $\theta'_{min}$, and 
$\theta'_{max}$ mentioned before and has an intrinsic origin, 
i.e. are related to the evolution of the active region and 
the structure of the jet/ejecta, there are two other effects that can create 
a lag: angular dependence of the emission and time/radius dependence of 
$\gamma_m$ and $\omega'_{cc}$. Assuming a homogeneous density for the 
colliding shells, equation (\ref{gammam}) shows that the time dependence of 
$\gamma_m$ is mainly due to variation of the relative Lorentz factor $\gamma'$ 
and the electric field energy fraction index $\epsilon_e$ with time. The 
variation of $\omega'_{cc}$ is also possible if the magnetic field changes 
with time. If fields are 
considered to be constant during the emission - as it is the case for many 
GRB models in the literature - only dynamical friction and decrease in 
$\gamma$ remain. They may be insufficient to explain the observed lags because 
the Doppler and high latitude correction is usually small, specially when 
$\Gamma$ is large. In fact simulations show that with an angular independent 
emission flux, large lags can be obtained only if the emission happens at 
large radius with respect to the central engine, $\gtrsim 10^{15}$ 
cm~\citep{dopplerlag1}. This is orders of magnitude larger than what is 
expected from internal shocks. As synchrotron emission is highly angular 
dependent, lags are expected not only from Doppler and high latitude 
correction but also when these effects are ignored. In fact this can be 
seen in (\ref{totpowinter}). The integration over $\gamma_e$ can be expressed 
as $f (\zeta_m)$ where $\zeta_m$ is $\zeta$ (defined in (\ref{zetadef})) for 
$\gamma_e = \gamma_m$. It is then clear that $r$ and $\omega'$ dependence of 
the integrand are not factorisable, and therefore even when ${\mathcal G}(r)$
is ignored, the $r$ dependence of $dP/\omega d\omega$ can not be factorized. 
Therefore, giving the expectation of high collimation and large Lorentz 
factor, the main cause of the observed lags in the GRBs seems to be the time 
variation of physical properties of the active region such as $\Delta r$, 
$\gamma_m$, and electric and magnetic fields.

We note that the zero-order term in (\ref{totpowinter}) is the same as the 
expression (\ref{synchpowp}) for the synchrotron emission by one particle. 
Therefore, for this part of integration in (\ref{totpowinter}) we can use 
the well known expression for the spectrum~\citep{synchspec,synchspec0,
electrody,radproc}. For determining the Doppler and high latitude correction 
term we can again use the integration method used for the zero-order term 
and express (\ref{totpowinter}) as a sum of Bessel 
functions~\citep{synchspec0}. 
Finally we find the power spectrum with the first order correction for 
Doppler effect and high latitude delays\footnote{For simplifying the 
integration over $\theta'$ in (\ref{totpowinter}), we consider 
$\cos \theta' + \beta \approx 2$. This is a valid approximation when 
$\beta \rightarrow 1$ and $\theta' \approx 0$ dominates the synchrotron 
emission.}:
\bea
\frac {dP}{\omega d\omega} & = & \frac{\sqrt {3} e^2}{3\pi} r^2 
\frac{\Delta r}{\Gamma (r)} \int_{\gamma_m}^\infty d\gamma_e n'_e 
(\gamma_e)\gamma_e^{-2} \nonumber \\
&& \biggl \{2 \int_{\frac{\omega'}{\omega'_c}}^{\infty} K_{5/3} (\zeta) 
d\zeta + \frac{7}{12}~{\mathcal G}(r) \biggl (\frac{\omega'_{cc}}
{\omega'}\biggr ) \nonumber \\
&& \biggl [ \frac{22}{7} K_{1/3} (\frac{\omega'}{\omega'_c}) - 9 
\biggl (\frac{\omega'}{\omega'_c}\biggr ) K_{2/3} (\frac{\omega'}
{\omega'_c}) + \nonumber \\
&& 3 \biggl (\frac{\omega'}{\omega'_c}\biggr ) 
\int_{\frac{\omega'}{\omega'_c}}^{\infty} K_{5/3} (\zeta) d\zeta \biggr ] 
\biggr \} \label{powerdopcorr}
\eea 
We can further integrate (\ref{powerdopcorr}) if we consider a specific 
distribution for accelerated electrons $n'_e (\gamma_e)$. In Appendix 
\ref{app:b} we calculate $dP/\omega d\omega$ for the power-law distribution 
(\ref{nedistpow}) and comment the case for a power-law with an exponential 
cutoff distribution. Here we use those results to discuss the lowest order 
properties of the spectrum and light curves. 

\subsection {Lag between light curves} \label {sec:lags}
The presence of a lag between various energy bands of the GRBs has been 
detected in BATSE~\citep{batse} and INTEGRAL\citep{integral} light curves and 
confirmed by \swift-BAT~\citep{batcat}. Observations show that in most long 
busts there is a significant lag - from tens to few hundreds of milliseconds 
between soft and hard bands. By contrast, in short bursts the lags is very 
small and within the present sensitivity and time resolution of gamma-ray 
telescopes it is consistent with zero or at most a few milliseconds. These 
two classes are respectively associated with the explosion of massive stars 
(collapsars, hypernova) and to the merging of two compact objects - a neutron 
star with a black hole, two neutron stars, or a neutron star and a white 
dwarf. There is however evidence for the difference between lags of the 
separate peaks in the same burst. Moreover, some bursts that according to 
their $T_{90}$ can be classified as short such as GRB 080426~\citep{grb080426} 
have relatively long lags of few tens of milliseconds. On the other hand some 
apparently long bursts such as GRB 060614~\citep{grb060614} and 
GRB 080503~\citep{grb080503} have small lags similar to the short bursts. 
Various explanation have been put forward for these {\it out of norm} 
behaviours: sensitivity of detractors only to the peak of a long burst 
leading to its misclassification as short, existence of a separate class of 
GRBs with intermediate durations and lags, long tail emission 
in otherwise short burst for bursts with long $T_{90}$ and small lags. Some 
authors even rule out the association of long GRBs - hypernova, short GRBs - 
collision of compact objects, and suggest that they should be classified 
according to their lags: Short lags old population, long lags young 
population~\citep{grbnewclass}. Apart from classification of progenitors of 
the GRBs, lags along with luminosity have been also used as proxy for the 
GRBs redshift determination~\citep{lagredshift}. 

In summary, lags seem to be important quantities related to the nature of 
central engine of the GRBs, properties of the ejecta, and the surrounding 
material. They can be relatively easily measured, and therefore it is 
important to be able to relate them to these phenomena. Some authors have 
tried to explain lags just as a geometrical effect related to the high 
latitude emission and associated Doppler 
shift~\citep{dopplerlag,dopplerlag0,dopplerlag1}. As we have discussed in the 
previous section and also regarding $r$ and $\omega$ dependence of the 
spectrum in (\ref{powerdopcorr}), it is evident that even without Doppler 
shift and high latitude corrections the $r$ dependence - equivalent to time 
dependence in the model discussed here - and $\omega$ dependence are not 
factorisable, and therefore light curves in different energy band can not be 
the same even when they are normalized to smear the amplitude difference.

To define and determine lags we need fast varying features such as peaks. In 
the gamma-ray bursts light curves peaks are mostly observed in the prompt 
gamma-ray energy bands. Nonetheless, fast slew of the \swift satellite and 
some of the ground based robotic telescopes have permitted to observe the 
counterpart peaks or at least evidence of their presence in X-ray and optical 
bands. A realistic model for the lags should be able to predict the lag 
in all these energy bands if they have the same origin. On the other hand, a 
deviation of some bands from predictions can be used as an evidence for a 
different origin of the corresponding feature. 

The nature of peaks and their profile are not well understood. In the 
framework of the synchrotron emission from internal shock model as the origin 
of the prompt gamma-ray emission, the rising side of a peak indicates the 
beginning of the collision between shells and formation of the electric and 
magnetic fields that leads to the acceleration of charged particles and 
synchrotron emission in the induced magnetic field. Decreasing edge 
corresponds to the separation and/or the total coalescence of the shells. 
However, it is expected that even before separation/coalescence, 
microphysics in the active region arrive to a roughly steady state during 
which only slight changes due to e.g. density fluctuation in the shells will 
occur. In particular when the initial evolution of the microphysics is much 
faster than the time of the passage of the shells through each other we 
expect that for a limited duration the active region has quasi steady 
characteristics. Assuming such a case - in accordance with the discussion 
about the evolution of $\Delta r$ in Sec. \ref{sec:shockevol} - a peak 
corresponds to:
\be
\frac {d}{dt} \biggl (\frac {dP}{\omega d\omega} \biggr ) = 
c \beta (r) \frac {d}{dr} \biggl (\frac {dP}{\omega d\omega} \biggr ) = 0 
\label{peakdef}
\ee
The lag for a given peak corresponds to the difference between peak 
time/radius for two frequencies or two energy bands. Usually observations 
are performed in known energy bands. Therefore, the purpose of the lag 
measurement is to determine the difference between $r_{peak}$ the solution of 
equation (\ref{peakdef}), at two different energies. In the rest of this 
section we use the results of Sec. \ref{sec:synchflux} to determine the lags.

Using (\ref{powerdopcorr}) the peak equation (\ref{peakdef}) can be written 
as a partial differential equation:
\bea
& & {\mathcal K}(r)\frac {dP}{\omega d\omega} + D(r) \frac{\partial}
{\partial \gamma_m^2}\biggl (\frac {dP}{\omega d\omega} \biggr ) + 
{\mathcal G}'(r) \frac{\partial}{\partial {\mathcal G}} \biggl (\frac {dP}
{\omega d\omega} \biggr ) = 0 \nonumber \\
\label {peakeq} \\
& & {\mathcal K}(r,\omega') \equiv \frac{1}{r^2 \Delta r}\frac{d}{dr}(r^2 
\Delta r) + \Gamma (r) \frac{d}{dr}\biggl (\frac {1}{\Gamma (r)} \biggr ) 
\label {kcoeff}\\
& & D (r) \equiv \frac{d\gamma_m}{dr} \label {dr} 
\eea
Note that the functions $D$ and ${\mathcal G}'$ depend only 
on $r$ and not on $\omega'$. If $\Delta r$ is energy independent, so is 
${\mathcal K}(r)$. The spectrum in (\ref{powerdopcorr}) can be written as:
\bea
\frac {dP}{\omega d\omega} & = & {\mathcal F}(r, \omega') 
\int_{\gamma_m}^\infty d\gamma_e n'_e (\gamma_e)\gamma_e^{-2} 
{\mathcal H}(r, \omega') \label {powh} \\
{\mathcal F}(r, \omega') & \equiv & \frac{\frac{\sqrt {3} e^2}{3\pi} r^2 
\Delta r}{\Gamma^4 (r)~(1-\beta (r))^3} \label {fdef} \\
{\mathcal H}(r, \omega') & \equiv & \biggl \{2 \int_{\frac{\omega'}
{\omega'_c}}^{\infty} K_{5/3} (\zeta) d\zeta + \frac{7}{12}~{\mathcal G}(r) 
\biggl (\frac{\omega'_{cc}}{\omega'}\biggr ) \nonumber \\
&& \biggl [ \frac{22}{7} K_{1/3} (\frac{\omega'}{\omega'_c}) - 9 
\biggl (\frac{\omega'}{\omega'_c}\biggr ) K_{2/3} (\frac{\omega'}
{\omega'_c}) + \nonumber \\
& & 3 \biggl (\frac{\omega'}{\omega'_c}\biggr ) \int_{\frac{\omega'}
{\omega'_c}}^{\infty} K_{5/3} (\zeta) d\zeta \biggr ] \biggr \} 
\label {hdef}
\eea
With this definition the partial differentials of $dP/\omega d\omega$ in 
the peak condition equation (\ref{peakeq}) can be calculated:
\bea
&& \frac{\partial}{\partial \gamma_m}\biggl (\frac {dP}{\omega d\omega} 
\biggr ) = -\gamma_m^{-2} n'(\gamma_m) {\mathcal F}(r, \omega') 
{\mathcal H}(r, \omega') \label {gammapar} \\
& & \frac{\partial}{\partial {\mathcal G}} \biggl (\frac {dP}{\omega d\omega} 
\biggr ) = {\mathcal F}(r, \omega') \int_{\gamma_m}^{\infty} d\gamma_e 
n'_e (\gamma_e) \gamma_e^{-2} \frac{\partial^2{\mathcal H}}{\partial \omega' 
\partial {\mathcal G}} \label{gpar} \\
& & {\mathcal K}(r,\omega') \int_{\gamma_m}^{\infty}n'_e (\gamma_e) 
\gamma_e^{-2}{\mathcal H} - n'_e (\gamma_m)\gamma_m^{-2} D(r){\mathcal H} + 
\nonumber \\
&& \quad\quad {\mathcal G}'\int_{\gamma_m}^{\infty}n'_e (\gamma_e) 
\gamma_e^{-2} \frac{\partial {\mathcal H}}{\partial {\mathcal G}} = 0 
\label {heq}
\eea
This equation is obviously very complex and solving it is not a trivial 
task. However, we are only interested in the change in the roots with respect 
to energy $\omega'$. Moreover, if we restrict ourselves to 
$|\Delta \omega'/ \omega'_0| \equiv |\omega'_1 / \omega'_0 - 1| < 1$ and 
assume that in this case the corresponding roots $r_0$ and $r_1$ are close 
i.e. $|r_1/r_0 -1| < 1$, then we can expand functions in (\ref{heq}) around 
$\omega'_0$ and determine the lag in the observer frame i.e. 
$c \beta \Delta t = r_1 - r_0$. Assuming that $\beta \approx 1$ we find:
\bea
c \Delta t & \approx & -\frac{\Delta\omega \frac {\partial^2}{\partial \omega 
\partial r} (\frac {dP}{\omega d\omega})}{\frac {\partial^2}
{\partial r^2} (\frac {dP}{\omega d\omega} )} = -\frac{\Delta\omega' \frac 
{\partial^2}{\partial \omega' \partial r} (\frac {dP}{\omega d\omega})}
{\frac {\partial^2}{\partial r^2} (\frac {dP}{\omega d\omega} )} = \nonumber \\
&& - \frac{\Delta \omega'}{\omega'} \frac {\omega' P(r_0, \omega'_0)}
{Q (r_0, \omega'_0)} \label {lag}\\
P(r_0, \omega'_0) & \equiv & {\mathcal K}(r_0) \int_{\gamma_m}^{\infty} 
d\gamma_e n'_e (\gamma_e) \gamma_e^{-2} \frac {\partial {\mathcal H}}
{\partial \omega'} - \nonumber \\
&& D (r_0) \frac {\partial}{\partial \omega'}(n'_e (\gamma_m) \gamma_m^{-2} 
{\mathcal H}) + {\mathcal G}'(r_0) \nonumber \\
&& \int_{\gamma_m}^{\infty} d\gamma_e n'_e (\gamma_e) \gamma_e^{-2} 
\frac{\partial^2{\mathcal H}}{\partial \omega' \partial {\mathcal G}} - 
\frac{1}{\omega' {\mathcal F}} \frac {d}{dr} \biggl (\frac {dP}{\omega 
d\omega} \biggr ) \nonumber \\
\label {lagp}\\
Q(r_0, \omega'_0) & \equiv & \frac {d{\mathcal K}}{dr}(r_0) \int_{\gamma_m}^
{\infty}n'_e (\gamma_e) \gamma_e^{-2}{\mathcal H} + {\mathcal K}(r_0) \biggl [
{\mathcal G}'(r_0) \nonumber \\
&& \int_{\gamma_m}^{\infty}n'_e (\gamma_e) \gamma_e^{-2} 
\frac {\partial {\mathcal H}}{\partial {\mathcal G}} - D n'_e (\gamma_m) 
\gamma_m^{-2}{\mathcal H} \biggr ] - \nonumber \\
& &\frac {\partial}{\partial r} (D n'_e (\gamma_m) \gamma_m^{-2}
{\mathcal H}) - {\mathcal G}'(r_0) D n'_e (\gamma_m) \gamma_m^{-2}
\frac {\partial {\mathcal H}}{\partial {\mathcal G}} + \nonumber \\
&& \frac{d{\mathcal G}'}{dr}(r_0) \int_{\gamma_m}^{\infty} n'_e (\gamma_e) 
\gamma_e^{-2} \frac{\partial {\mathcal H}}{\partial {\mathcal G}} + 
\frac {{\mathcal K}}{{\mathcal F}} \frac {d}{dr} \biggl (\frac {dP}
{\omega d\omega} \biggr )\label {lagq}
\eea
All the terms in (\ref{lagp}) and (\ref{lagq}) can be expressed as a sum of 
Bessel functions and their integrals using the definition of ${\mathcal H}$. 
For the special case of a power-law distribution of electrons most of the 
integrals can be calculated analytically. We present the results in 
Appendix \ref{app:b}. Note also that 
$\Delta \omega'/\omega' = \Delta \omega /\omega$ and 
$\omega'\partial/\partial \omega' = \omega \partial/\partial \omega$. Other 
$\omega$ dependent terms also are $\Gamma$-independent and therefore the 
calculation of the lag does not need a pre-knowledge of the bulk Lorentz 
factor $\Gamma$.

The left hand side of (\ref{peakdef}) or equivalently (\ref{peakeq}) is the 
slope of the spectrum (light curve) and therefore its energy dependence 
explains for instance why peaks or more precisely their auto-correlations are 
wider at lower energies~\citep{peakwidth}. However, the complexity of 
expressions (\ref{lag}) to (\ref{lagq}) make the analytical estimation of the 
time and energy dependence of light curves and lags difficult. In Paper II we 
present some simulations in which the lags are consistent with the long bursts 
or are very small similar to short bursts~\citep{shotgrbrev}.

\subsection {Break at low energies}\label{sec:break}
Before finishing this section we want to make a few comments about the jet 
break which is one of the most important aspects of the afterglow light 
curves predicted since the beginning of the modelling of GRBs. 

Since the early days of the discovery of GRBs and measurement of 
apparently huge amount of energy released in these phenomena - 
${\mathcal O}(0.1-10) \times 10^{53}$ ergs for long bursts or even larger in a 
few exceptionally bright bursts such as GRB 990123, 
GRB 080319B~\citep{grb080319b,grb080319b1} and GRB 080607~\citep{grb080607} - 
it has been suggested that these measurements are biased by the collimation of 
the fireball and the actual total emitted energy should be much smaller. To 
produce the gamma-ray prompt emission in a shock the ejecta should be highly 
relativistic with a bulk Lorentz factor of order of few hundreds. This leads 
to a strong apparent collimation of the radiation from a spherically 
symmetric ejecta to angles $\Theta < \Theta_{boost} \equiv 1/\Gamma$ for a far 
observer~\citep{relcollim}. It is also possible that the ejecta is not 
intrinsically spherical but jet like~\citep{grbag}. In this 
case after deceleration of the fireball during its propagation through the 
surrounding material or the ISM, at some radius/time 
$\Theta_{boost} > \theta_{open} = \theta_{max} - \theta_{min}$ and radiation 
is no longer collimated. This lead to a drop of observed 
flux~\citep{jetbreak}. 
As this effect is purely kinematic/geometric it should not depend on energy. 
Observations of the \swift and robotic ground telescopes however contradict 
this expectation. The breaks seen in the afterglow light curves are usually 
chromatic. In many bursts no optical break has been observed up to millions 
of seconds after the prompt emission~\citep{uvotlc}. In some bursts, mostly 
bright ones but not always, no break has been observed in X-ray 
either~\citep{nojetbreak}. In a significant fraction of bursts multiple 
breaks in the X-ray light curve have been observed. Apriori only one of these 
breaks (if any) can be due to the jet break. Therefore, the mechanism of break 
is more complex than just a kinematical effect. Here we want to argue that 
even in the simple case of the opening due to deceleration, we should not 
expect to have an achromatic break if the afterglow is a synchrotron radiation.

For obtaining the analytical expression (\ref{powerdopcorr}) for the radiation 
spectrum, we made the simplifying approximation that $|\theta_{min}| = 
|\theta_{max}| \rightarrow \infty$. This permitted to analytically integrate 
the angular integral in (\ref{totpowinter}). Without this approximation 
(\ref{totpowinter}) depends on the opening angle. The justification for this 
approximation was that the synchrotron emission is highly directional and for 
$\theta' \gamma_e \gg 1$ the intensity reduces exponentially because the value 
of $\zeta$ in the integrand of (\ref{totpowinter}) grows. $\zeta$ increases 
with energy and therefore high energy photons are preferentially emitted at 
small angles with respect to electrons boost direction. When the collimation 
is reduced, the observer receives less high energy photons from high latitude 
electrons~\citep{ringemission} and therefore the effect of the jet break 
influences high energy light curves earlier than low energy ones. 
Conceptually this effect is very similar to the lag between light curves. It 
is however more difficult to investigate it mathematically because the 
angular integral (\ref{totpowinter}) can not be determined analytically as 
explained above.

On the other hand, it is not sure that the reduction of Lorentz factor can 
explain the absence of a break in optical frequencies when it is observed in 
X-ray. A number of suggestions have been put forward to explain 
observations~\citep{jetbreakexp,jetbreakexp0}. In the frame work of the model 
presented here it can be at least partly explained by the energy dependence 
of the geometry of the active region. In fact the material that has lost its 
energy and has been decelerated is pushed behind 
the front edge of the shock. Due to the scattering it also has a relatively 
larger transverse momentum (See simulations in~\citep{pulsarbowshock} for 
non-relativistic shocks). Thus, less boosted particles in the ejecta have 
intrinsically a larger opening angle. This leads to a later jet break for 
the softer photons originating in this extended region. Therefore one does 
not need additional processes~\citep{intext1,refresh1}, 
energy~\citep{enerinj,enerinj1,enerinj2} and/or 
components~\citep{jetbreakexp1} to explain the breaks. The complex geometry 
of matter and fields in the material and shock can explain apparently 
contradictory observations. As for multiple breaks, they can be due to the 
fact that multiple shells are ejected by the central engine. This is in the 
same spirit as the structured jet models~\citep{structedjet,structedjet0}. 
If they do not completely coalesce both the tail emission and the afterglow 
due to the external 
shocks will combinations of radiation from separate remnant shells. In this 
case the break of radiation from each shell is independent of others and a 
far observer who detects the total emission will observe multiple breaks in 
the emission.

\section {Extraction of parameters from data} \label{sec:extract}
The model described in the previous sections has a large number of parameters. 
It would not help to better understand the physics of gamma-ray bursts 
production and their engine if we can not estimate the parameters of this 
or any other model from data. On the other hand, the only conveyor of 
information for us 
is the emission in different energy bands. Therefore, we must find the best 
ways to extract the information as efficient as possible. In this section we 
suggest a procedure for extracting the parameters of the model. We also 
discuss their degeneracies.

Since the massive detection of GRBs by BATSE, many efforts have been 
concentrated to understand their spectrum. However, very little information 
could be extracted from the spectra. The complexity of time and energy 
dependence of the spectrum (\ref{powerdopcorr}) explains why the time averaged 
spectrum does not carry extractable information. As GRBs evolve very quickly 
both in time and in energy, integration over these quantities smears the 
useful information. Unfortunately the effective detection surface of present 
gamma-ray detectors is too small and we do not have enough photons to make a 
broad spectrum in small time intervals. In this situation hardness ratios 
and their time variation are more useful quantities. Therefore we first 
discuss what we can extract from hardness ratios. 

Consider a power-law distribution for electrons, equation (\ref{powerpowlaw}) 
shows that at a given time the spectrum can be expanded as a power of 
$\omega'/\omega'_{cc}\gamma_m^2$. The coefficient in front of the integral 
term does not depend on energy. Therefore, if we neglect $\mathcal G (r)$ i.e. 
assuming the bulk Lorentz factor is very large, the hardness ratios depend 
only on the integral determined in (\ref{firstkint}). We define the hardness 
ratio between two bands with logarithmic mean energies $\omega_1$ and 
$\omega_2$ as:
\bea
HR_{12}(r) &=& \frac {\int_{\omega^{min}_1}^{\omega^{max}_1} d\omega 
\frac {dP}{\omega d\omega}}{\int_{\omega^{min}_1}^{\omega^{max}_1} d\omega 
\frac {dP}{\omega d\omega}} \approx \frac {\frac {dP}{\omega d\omega}
\biggl |_{\omega = \omega_1}}{\frac {dP}{\omega d\omega}\biggl |_{\omega = 
\omega_1}} \times \frac{\Delta \omega_1}{\Delta \omega_2} \label{hardnessdef}\\
&& \Delta \omega_i \equiv \omega^{max}_i - \omega^{min}_i \label{banddef}
\eea
The approximations for integrals are valid when $\Delta \omega_i / \omega_i 
\ll 1$. Expression (\ref{firstkint}) shows that at a constant time/radius 
hardness ratio depends only on $p$, electron distribution index, and on the
Lorentz invariant factor $\omega' / \omega'_m$ where:
\be
\omega'_m \equiv \omega'_{cc}\gamma_m^2 \label {omegamdef}
\ee
is the minimum synchrotron characteristic frequency. Therefore by fitting 
observational data with (\ref{hardnessdef}) one can obtain the index of the 
electron distribution and the time variation of $\omega_m$ which is a very 
important quantity. Note that as $HR$ is time/radius dependent we need 
multiple energy bands and their hardness ratios to make a statistically 
meaningful fit. If we release the assumption of power-law distribution for 
electrons, we can use (\ref{powerdopcorr}) to get insight into the 
distribution of electrons Lorentz factor and $\omega_m$, but they will be 
degenerate.

Once the energy and temporal behaviour of the integral term in 
(\ref{powerdopcorr}) is found, the extraction of leading time dependent 
coefficient ${\mathcal F}(r)$ (defined in (\ref{fdef})) from light curves 
is easy. Then if we use one of the models considered in this work or another 
model for the evolution of $\Delta r$, we can determine time/radius 
variation of $\Gamma (r)$ the total Lorentz factor of the active 
region. Using (\ref{gammabulk}) and (\ref{betan}) one can apriori estimate 
$\mathcal M (r')$. However, the latter depends on the relative Lorentz 
factor $\gamma' (r')$. If we can roughly estimate the end of the collision 
from the form of the light curve - for instance if we assume that the end of 
the collision is when the continuous component begins an exponential 
decay - then:
\be
\Gamma_f = \Gamma (r) \biggl |_{r = r_f} \label{gammaf}
\ee
where $r_f$ is the radius at which the coalescence of the shells finishes.

$\omega' / \omega'_m$, $\mathcal M (r')$, and $\gamma_m (r')$ depend on the 
fundamental quantities such as magnetic field and thereby on the fraction of 
kinetic energy of the shell transferred to the electric and magnetic fields, 
the density of shells, and the initial distance of the collision from the 
central engine. But with only 3 quantities mentioned above we can not 
determine all these quantities and their time/radius variation. An important 
observable that can help to extract more information from data is the lag. 
If we neglect the term proportional to $\mathcal G (r)$ and its derivatives, 
we can determine functions $P (r_0, \omega')$ and $Q (r_0, \omega')$ in 
(\ref{lagp}) and (\ref{lagq}) from the knowledge about ${\mathcal F}(r)$ and 
the integral of ${\mathcal H}(r)$ (defined in (\ref{hdef})) explained above. 
Thus in this way we can determine the lag between two energy bands. Comparison 
of the observed lags with the expectations from the model helps to estimate 
some of other quantities such as $r_0$. However, as the analytical 
expression for the lags is very complex, only simulations and fit can 
permit to solve this inverse problem. Using a simpler version of the model 
presented here, this procedure has been applied to analyse the \swift data 
for GRB 060607a~\citep{grb060607a} and to explain some of its peculiar properties  
~\citep{grb060607a1}.

At the beginning of this section we mentioned that the most complete 
information about GRBs is in the spectrum at short time intervals when the 
flux does not change significantly. In fact in practice hardness ratios are 
calculated by adding together photons in a given energy band and time-rebinned 
event data to reduce the noise. When multiple simultaneous hardness ratios 
are available, they can be considered as a low resolution normalized spectrum 
in a short time interval. Because these quantities give the most direct 
insight into the physics of the collision, in the rest of this section we 
investigate in more details the spectral behaviour of the flux at constant 
time. 

First we consider a power-law distribution for electrons. When 
$\omega'_m > \omega'$ hypergeometric functions$~_1F_2$ in (\ref{firstkint}) 
and (\ref{secondkint}) can be expanded as a polynomial with positive or zero 
power of $\omega'/\omega'_m$. This shows that at the lowest order in 
$\omega'/\omega'_m$ the spectrum $dP/\omega d\omega \propto 
(\omega'/\omega'_{m})^{-2/3}$ and therefore $HR_{12} \propto 
(\omega_1/\omega_2)^{-2/3}$. The power of the coefficient in the 
second$~_1F_2$ term is larger and therefore for determining the spectrum at 
zero order it can be neglected. But for the first order expansion of 
(\ref{powerpowlaw}) this term is significant and when it is added the spectrum 
becomes:
\be
\frac {dP}{\omega d\omega} \propto \biggl [(\omega'/\omega'_m)^{-2/3} + 
\frac{\frac{p}{4} + \frac{1}{6}}{2^{\frac{4}{3}} (\frac{p}{4} + 
\frac{5}{6})}(\omega'/\omega'_m)^{2/3} \biggr ] \quad \mbox{for }
\frac {\omega'}{\omega'_m} < 1 \label {ppowlawfirst}
\ee
The amount of deviation from the dominant power-law only marginally depends 
on the index of the electron distribution function $p$. When 
$\omega'/\omega'_m \rightarrow 1$ the higher power of $\omega'/\omega'_m$ 
become important and should be considered. However, as they are all 
positive, they make the spectrum harder. This result shows that 
although small hardening of the spectrum can be obtained in this regime, 
it is not possible to have a softer spectrum with an index smaller than 
$-2/3$ and closer to what has been observed in many 
bursts~\citep{batcat,shotgrbrev}. Note also that the addition of terms 
proportional to ${\mathcal G}(r)$ can not make the spectrum softer because 
they have the same form of energy dependence as the dominant terms.

Next we consider the regime where $\omega'/\omega'_m > 1$. In this regime 
there is no analytical expression for~$_1F_2$ functions. Therefore we use the 
asymptotic behaviour of Bessel functions, $K_\nu (x) \approx \sqrt {\pi/2x} 
e^{-x}$, to estimate the integrals in (\ref{powerpowlaw}). Integration of the 
left hand side of (\ref{firstkint}) leads to:
\bea
& &\int_{\gamma_m}^\infty d\gamma_e \biggl (\frac{\gamma_e}{\gamma_m} 
\biggr )^{-(p+1)}\gamma_e^{-2} \int_{\frac{\omega'}{\omega'_c}}^{\infty} 
K_{5/3} (\zeta) d\zeta \approx \nonumber \\
&& \frac{1}{\gamma_m} \biggl \{\sqrt{\frac{\pi}{2}}
(\frac{\omega'}{\omega'_m})^{\frac{1}{2}} \biggl [\Gamma (-(p+3), 
(\frac{\omega'}{\omega'_m})^{\frac{1}{2}}) - \Gamma (-(p+3), 1)\biggr ] + 
\nonumber \\
& & (\frac{\omega'}{\omega'_m})^
{-\frac{p+2}{2}} \biggl [2^{-\frac{1}{3}} \frac{\Gamma(\frac{2}{3})}
{\frac {p}{4} + \frac{1}{6}}~_1F_2 (\frac {p}{4} + \frac{1}{6}; \frac{1}{3}, 
\frac {p}{4} + \frac{7}{6}; \frac{1}{4}) + \nonumber \\
& & 2^{-\frac{5}{3}} \frac{\Gamma(-\frac{2}{3})}{\frac {p}{4} + 
\frac{5}{6}}~_1F_2 (\frac {p}{4} + \frac{5}{6}; \frac{5}{3}, \frac {p}{4} + 
\frac{11}{6}; \frac{1}{4})\biggr ]\biggr \} \quad \mbox{for }
\frac {\omega'}{\omega'_m} > 1 \label{ppowlawsec}
\eea
The constant coefficient in the last term is $\sim 1.354 (2^{5/3}/(p+2/3) + 
2^{1/3}/(p + 10/3))$. Using an expansion in series of the incomplete $\Gamma$ 
function:
\be
\Gamma (a, x) = e^{-x} x^a \biggl [1+\frac{1}{x + \frac{1-a}{1 + \frac{2-a}
{x + \ldots}}} \biggr ] \label{gammaseries}
\ee
we see that in this regime at the lowest order the spectrum is a power-law 
with an index $\sim -(1+p/2)$ which is much steeper (softer) than when 
$\omega'/\omega'_m < 1$. For energies between these two extremes regimes one 
expect an index $-(1+p/2) \lesssim \alpha \lesssim -2/3$. They can be fitted 
by a power-law, but at high $\omega'/\omega'_m$ a power-law with an 
exponential cutoff gives a slightly better fit due to the presence of the 
exponential term in the expansion of the incomplete $\Gamma$ functions. 
Nonetheless if the distribution of electrons is truncated at high Lorentz 
factors or is a broken power-law in which the index becomes smaller at higher 
energies, this constraint will not be applicable. As $\omega'/\omega'_m > 1$, 
the second term in the sum and the last term in (\ref{ppowlawsec}) are smaller 
than other terms considered above.

The two regimes explained above cover the maximum and soft wing of the index 
range observed by \swift~\citep{batcat} and BATSE but it can not explain very 
hard short bursts with indices larger than the zero order index $-2/3$. 
Moreover, when the observed spectrum is extended to high energies, 
$\gtrsim 1 MeV$, a power-law with an exponential cutoff at high energies 
is a better fit to GRBs spectrum. This type of spectra can be obtained if 
the distribution of the Lorentz factor of the accelerated electrons has an 
exponential cutoff at high energies. As we mentioned in Sec.\ref{sec:shockevol}
the spectrum of the accelerated electrons in the simulations of 
ultra-relativistic shocks is more sophisticated than a simple power-law and is 
closer to a power-law with an exponential cutoff~\citep{fermiaccspec}, 
see (\ref{fermispec}). Assuming for simplicity an exact power-law with 
exponential cutoff for the Lorentz factor distribution of accelerated 
electrons like eq. (\ref{nedistpowc}), the dominant term of the spectrum 
(\ref{powerdopcorr}) becomes proportional to:
\be
\int_{\gamma_m}^\infty d\gamma_e \biggl (\frac{\gamma_e}{\gamma_m} \biggr )^
{-(p+1)} e^{-\frac{\gamma_e}{\gamma_p}}\gamma_e^{-2} \int_{\frac{\omega'}
{\omega'_c}}^{\infty} K_{5/3} (\zeta) d\zeta \label{plcutint}
\ee
with the conditions mentioned in (\ref{nedistpowc}). When $\gamma_e \ll 
\gamma_{cut}$ the exponential term is close to $1$ and the behaviour of the 
spectrum is indistinguishable from a simple power-law. But at high energies 
photons are emitted preferentially by faster electrons for which the 
exponential term is important and leads to an exponential cutoff in the 
spectrum of the synchrotron emission. To prove such a behaviour, we 
concentrate on the part of the spectrum for which 
$\omega'/\omega'_m \gtrsim 1$. In this case the integration of 
(\ref{plcutint}) leads to:
\bea
&& \int_{\gamma_m}^\infty d\gamma_e \biggl (\frac{\gamma_e}{\gamma_m} \biggr )^
{-(p+1)} e^{-\frac{\gamma_e}{\gamma_p}}\gamma_e^{-2} \int_{\frac{\omega'}
{\omega'_c}}^{\infty} K_{5/3} (\zeta) d\zeta \approx \nonumber \\
&& \frac{\sqrt {\frac{\pi}{2}}}{\gamma_m} \biggl [
\int_1^{(\frac{\omega'}{\omega'_m})^{\frac{1}{2}}} dy y^{-(p+3)} 
e^{-\frac{\gamma_m}{\gamma_p} y} \Gamma (\frac{1}{2}, \frac{\omega'}
{\omega'_m y^2}) + \nonumber \\
&& 2 \int_{(\frac{\omega'}{\omega'_m})^{\frac{1}{2}}}^
\infty dy y^{-(p+3)} e^{-\frac{\gamma_m}{\gamma_p} y} K_{2/3} (\frac{\omega'}
{\omega'_m y^2})\biggr ] \label {powerexpodist}
\eea
For $\omega'/\omega'_m \gg 1$ the second integral on the right hand side of 
(\ref{powerexpodist}) is small and negligible. The first integral does not 
have analytical solution, but using asymptotic behaviour of incomplete 
$\Gamma$ function (\ref{gammaseries}) one can conclude that its behaviour at 
high energies must be close to an exponential with negative exponent 
proportional to $\omega'$.

In the beginning of this section we mentioned that most probably $\Delta r$ 
is also somehow energy dependent, specially at very low and very high 
energies. In the latter case it must be very small, probably exponentially 
decreasing with energy. This can also contribute to the existence of an 
exponential cutoff at high energies. As for the low energy tail of the 
spectrum the fact that even the residual energy of cooled electrons can be 
enough to emit soft photons means that the spectrum in this regime should be 
much flatter. This is consistent with the roughly flat spectrum and shallow 
temporal decline in optical and longer wavelength emissions as observed in 
most GRBs~\citep{uvotlc,sampaper}.

In conclusion of this section, we showed that when a relativistic shock 
is modelled in details and realistic distributions for electrons are 
considered, the synchrotron emission alone can explain different aspects of 
the time averaged spectrum of GRBs as well as their spectrum in a short 
interval in which the evolution can be ignored. Evidently, other processes 
such as inverse Compton~\citep{comptonreverse,comptontev} and pair 
production~\citep{grbtev} contribute to the total emission but most probably 
are not the dominant component at low/intermediate energy bands. Nonetheless 
their contribution should be more important at GeV and higher 
energies~\citep{grbtev}.

\section{Summary}
We presented a formulation of the relativistic shocks and synchrotron emission 
that includes more details than the dominant term considered in the previous 
calculations. Although we consider spherical shells, most of our results are 
valid also for non-spherical collimated jets as long as the collimation angle 
due to the relativistic boost is smaller than collimation angle. 

We showed that the lags between light curves at different energies exist in 
the dominant order and are not only due to the high latitude emissions which 
are negligible for ultra-relativistic ejecta.  The main reason for such 
behaviour is the evolution of electric and magnetic fields as well as the 
evolution of the emitting region which can be in addition energy dependent. 
This fact is more evident in the simulations presented in 
Paper II. Despite the absence of high latitude terms in our simulations, the 
presence of lags between the light curves of different energy bands is 
evident. For the same reasons the change in the slope of the light 
curves - what is called breaks - are also energy dependent. This explains 
chromatic breaks of the GRBs detected by \swift.

The two phenomenological models we considered for the evolution of the active 
region are physically motivated, but do not have rigorous support from 
microphysics of the shock. Nonetheless, they can be easily replaced if 
future simulations of Fermi processes lead to a better estimation of the 
size of the region in which electric and magnetic fields are formed and 
particles are accelerated and dissipated. Presence of an external magnetic 
field in the environment of the candidates for central engine of GRBs is very 
plausible. The formulation present here does not include such a possibility, 
but an external magnetic field can be added to (\ref{magener}). The 
modification of the evolution equation of $\beta'$ and the flux is 
straightforward.

Many other details such as the effect of metalicity of the ejecta and 
surrounding material both on the low energy emission and absorption is not 
considered in this work. We have also neglected synchrotron self-abortion. 
It only affects the low energy bands, nonetheless in hard bursts even 
optical emission can be affected by self absorption. We leave the study of 
this issue, the effect of ionization on the emissions, and the thermalization 
of shocked material to future works. 

\appendix
\section {Analytical approximation of $\beta'_{(1)}$} \label {app:a}
The integral in (\ref{integdyn}) can not be taken analytically. An analytical 
approximation is however useful for investigating the effect and importance 
of the radiation in the dynamics of the shell collision. Here we consider a 
few approximations for various range of ${\mathcal D}$ and $\eta$ that 
determine respectively the strength of the shock and the radiation.

If $\eta = \tau = 0$, the integral in (\ref{integdyn}) can be calculated 
analytically:
\bea
&& \int_{\beta'_0}^{\beta'_{(0)} (r')} dy~\frac{1}{y^3 (1-y^2)^3} = 
\biggl [A_1\ln y + \frac{A_2}{y} + \frac{A_3}{2y^2} + \nonumber \\
&& \quad B_1\ln (y -1) + \frac{B_2}{y - 1} + \frac{B_3}{2(y - 1)^2} + 
C_1\ln (y + 1) + \nonumber \\
&& \quad \frac{C_2}{y + 1} + \frac{C_3}{2(y + 1)^2} 
\biggr ]_{\beta'_0}^{\beta'_{(0)} (r')} \label{etatauzero}
\eea
\bea
&& A_1 = 3, \quad A_2 = 0, \quad A_3 = 1, \quad B_1 = \frac{11}{8}, \quad 
B_2 = \frac{1}{2}, \nonumber \\
&& B_3 = \frac {1}{8}, \quad C_1 = \frac{9}{16}, \quad C_2 = -\frac {9}{16}, 
\quad C_3 = -\frac{1}{8}. \label{etataucoeff}
\eea
When $\eta /(3-\kappa)$ is large, as the value of $y$ in the integrand of 
(\ref{integdyn}) is always less than $1$ we can formally expand the term 
$(1 - y^2)^{3-\tau/2}$ and integrate term by term:
\bea
& & {\mathcal M}_{(0)}(r') = \frac {\Delta r'_0}{(3-\kappa)} \biggl (
\frac{\gamma'_0}{\beta'_0} \biggr )^{\tau} ((3-\kappa) {\mathcal D})^{1-
\frac{\eta}{3-\kappa}} \biggl \{ \frac {1}{\tau - 2 + \frac{\eta}{3-\kappa}} 
\nonumber \\
& & \quad \bigg [{\beta'_0}^{\tau - 2 + \frac{\eta}{3-\kappa}}~_2F_1 
[\frac {\eta}{3-\kappa}, \tau - 2 + \frac{\eta}{3-\kappa}; \tau - 1 + 
\frac{\eta}{3-\kappa}; \nonumber \\
& & \quad 1 - \frac {\beta'_0}{(3-\kappa) {\mathcal D}}] - {\beta'_{(0)}}^
{\tau - 2 + \frac{\eta}{3-\kappa}}~_2F_1 [\frac {\eta}{3-\kappa}, \tau - 2 + 
\frac{\eta}{3-\kappa};\nonumber \\
& & \quad \tau - 1 + \frac{\eta}{3-\kappa}; \frac{\beta'_{(0)}(r')}{\beta'_0}
(1 - \frac {\beta'_0}{(3-\kappa) {\mathcal D}})]\biggr ] + \frac {(3 -
\frac{\tau}{2})}{\tau + \frac{\eta}{3-\kappa}} \nonumber \\
& & \quad \biggl [{\beta'_0}^{\tau + \frac{\eta}{3-\kappa}}~_2F_1 [\frac 
{\eta}{3-\kappa}, \tau + \frac{\eta}{3-\kappa}; \tau + 1 + 
\frac{\eta}{3-\kappa}; \nonumber \\
& & \quad 1 - \frac {\beta'_0}{(3-\kappa) {\mathcal D}}] - {\beta'_{(0)}}^
{\tau + \frac{\eta}{3-\kappa}}~_2F_1 [\frac {\eta}{3-\kappa}, \tau + 
\frac{\eta}{3 - \kappa}; \nonumber \\
& & \quad \tau + 1 + \frac{\eta}
{3-\kappa}; \frac {\beta'_{(0)}(r')}{\beta'_0}(1 - \frac {\beta'_0}{(3-\kappa) 
{\mathcal D}})]\biggr ] + \ldots \biggr \} \label {dynmzero}
\eea
Here$~_2F_1$ is the hypergeometric function and can be expanded as a 
polynomial of its last argument. When the latter is less than $1$ and the 
power of the terms in the polynomial are positive, they converge rapidly to 
zero. When the last argument in$~_2F_1$ is larger than one an analytical 
extension of this function with negative power in the polynomial expansion 
exists~\citep{integbook}. Therefore, the dominant term is always 
$\beta'_{(0)}$ in front of each $_2F_1$ term. For small $\eta$ this 
approximation is not valid. In this case we expand 
$(1 + (\frac {1}{(3-\kappa) {\mathcal D}} - 
\frac {1}{\beta'_0}) y)^{-\frac{\eta}{3-\kappa}}$ and obtain:
\bea
& & {\mathcal M}_{(0)}(r') = \frac {\Delta r'_0}{(3-\kappa)} \biggl (
\frac{\gamma'_0}{\beta'_0} \biggr )^{\tau} [(3-\kappa)\mathcal D]^{1-
\frac{\eta}{3-\kappa}} \biggl \{\frac{1}{\tau - 2 + \frac{\eta}{3-\kappa}} 
\nonumber \\
& & \quad \biggl [{\beta'_0}^{\tau - 2 + \frac{\eta}{3-\kappa}}~_2F_1 (3 - 
\frac{\tau}{2}, \frac{\tau}{2} - 1 + \frac{\eta}{2(3-\kappa)}, 
\frac{\tau}{2} + \frac{\eta}{2(3-\kappa)}; \nonumber \\
& & \quad {\beta'_0}^2) - {\beta'_{(0)}}^{\tau - 2 + \frac{\eta}{3 - 
\kappa}}~_2F_1 (3-\frac{\tau}{2}, \frac{\tau}{2}-1 + \frac{\eta}{2(3-\kappa)}, 
\nonumber \\
& & \quad \frac{\tau}{2} + \frac{\eta}{2(3-\kappa)}; {\beta'_{(0)}}^2(r')) 
\biggr ] - \frac {\frac{\eta}{3-\kappa} [\frac{1}{(3-\kappa){\mathcal D}} - 
\frac{1}{\beta'_0}]}{\tau - 1 + \frac{\eta}{3-\kappa}} \nonumber \\
& & \quad \biggl [{\beta'_0}^{\tau - 1 + \frac{\eta}{3-\kappa}}~_2F_1 (3 - 
\frac{\tau}{2}, \frac {\tau}{2} - \frac{1}{2} + \frac{\eta}{2(3-\kappa)}; 
\frac {\tau}{2} + \frac{1}{2} + \nonumber \\
& & \quad \frac{\eta}{2(3-\kappa)}; {\beta'_0}^2) - {\beta'_{(0)}}^{\tau - 1 + 
\frac{\eta}{3-\kappa}}~_2F_1 (3-\frac{\tau}{2}, \frac{\tau}{2} - 
\frac {1}{2} + \nonumber \\
& & \quad \frac{\eta}{2(3-\kappa)}; \frac{\tau}{2} + \frac{1}{2} + 
\frac{\eta}{2(3-\kappa)}; {\beta'_{(0)}}^2(r')) \biggr ] + \ldots \biggr \} 
\label {dynmzeroetazero}
\eea
In the same way we find the following expression for ${\mathcal M}_{(0)}(r')$ 
for the quasi steady model and large $\eta/(3-\kappa)$:
\bea
& & {\mathcal M}_{(0)}(r') = \frac {\Delta r_{\infty}}{3-\kappa} ((3-\kappa) 
{\mathcal D})^{1-\frac{\eta}{3-\kappa}} \biggl \{ \biggl \{ \frac {1}
{\frac{\eta}{3-\kappa} - 2}\bigg [{\beta'_0}^{\frac{\eta}{3-\kappa} - 2} 
\nonumber \\ 
& & \quad~_2F_1 [\frac {\eta}{3-\kappa}, \frac{\eta}{3-
\kappa} - 2; \frac{\eta}{3-\kappa} - 1; 1 - \frac {\beta'_0}{(3-\kappa) 
{\mathcal D}}] - \nonumber \\ 
& & \quad {\beta'_{(0)}}^{\frac{\eta}{3-\kappa} - 2}~_2F_1 [\frac 
{\eta}{(3-\kappa)}, \frac{\eta}{3-\kappa} - 2; \frac{\eta}{3-\kappa} - 1; 
\nonumber \\
& & \quad \frac {\beta'_{(0)}(r')}{\beta'_0}(1 - \frac {\beta'_0}{(3-\kappa) 
{\mathcal D}})]\biggr ] + \frac {1}{\eta} \biggl [{\beta'_0}^{\frac{\eta}{3-
\kappa}}\nonumber \\
& & \quad~_2F_1 [\frac {\eta}{3-\kappa}, \frac{\eta}{3-\kappa}; 1 + \frac{\eta}
{3-\kappa}; 1 - \frac {\beta'_0}{(3-\kappa) {\mathcal D}}] - \nonumber \\
& & \quad {\beta'_{(0)}}^{\frac{\eta}{3-\kappa}}~_2F_1 [\frac {\eta}
{3-\kappa}, \frac{\eta}{3-\kappa}; 1 + \frac{\eta}{3-\kappa}; \frac 
{\beta'_{(0)}(r')}{\beta'_0}(1 - \nonumber \\
& & \quad \frac {\beta'_0}{(3-\kappa) {\mathcal D}})] \biggr ] + 
\ldots \biggr \} - ((3-\kappa){\mathcal D})^{-\frac {\delta}{3-\kappa}} 
\nonumber \\
& & \quad \biggl \{\mbox {Same as above with } \eta \rightarrow \eta + 
\delta \biggr \} \biggr \} \label {steadymzero}
\eea
For the case of a small $\eta/(3-\kappa)$ one can use (\ref{dynmzeroetazero}) 
and make an expression analogue to (\ref{steadymzero}), thus we do not 
repeat the details here. We note that when $\tau = 0$ and $\delta \rightarrow 
\infty$, i.e. for a constant $\Delta r'$, (\ref{dynmzero}) and 
(\ref{steadymzero}) are equal as expected. When $\delta \rightarrow 0$, 
the formation of the active region is very slow. In this case the energy 
loss by radiation becomes negligible and $\beta$ decreases only due to the 
increasing accumulated mass. Note also that ${\mathcal M}_{(0)}(r_0) = 0$ as 
expected. If the active region varies according to (\ref{drquasiend}), the 
expression for ${\mathcal M}_{(0)}(r')$ includes only the term 
$\eta \rightarrow \eta + \delta$ in (\ref{steadymzero}) with a positive sign 
in front.

Although equations(\ref {dynmzero}), (\ref {steadymzero}), and 
(\ref{dynmzeroetazero}) 
seem quite sophisticated, due to the polynomial representation of $_2F_1$, 
only the few dominant terms are of real interest to us. In most cases we are 
only interested in the dominant power-law component. However, having 
expressions beyond the dominant power permits to go much further and 
calculate quantities such as lags that in a simple power-law approximation 
can not be determined. 

When the kinetic energy of the fast shell does not change significantly 
$\beta'_{(0)}(r') 
\sim \beta'_0$ and ${\mathcal M}_{(0)}(r) \rightarrow 0$. This case happens 
when the radiation has a negligible effect on the kinematic of the shock. 
Assuming a strong shock i.e. $\beta'_0 < 3{\mathcal D}$, we can use the 
definition of $~_2F_1$ to investigate the behaviour of ${\mathcal M}_{(0)}(r)$ 
at lowest order. For the dynamically driven active region with large $\eta$ 
and a relatively soft shock, when we expand$~_2F_1$ terms in (\ref {dynmzero}) 
up to first order, ${\mathcal M}_{(0)}(r)$ becomes:
\bea
&& {\mathcal M}_{(0)}(r') \approx \frac {\Delta r'_0}{(3-\kappa)} \biggl (
\frac{\gamma_0}{\beta'_0} \biggr )^{\tau} ((3-\kappa) {\mathcal D})^{1-
\frac{\eta}{3-\kappa}} {\beta'_0}^{\tau - 2 + \frac{\eta}{3-\kappa}} 
\nonumber \\
& & \quad \biggl \{\frac{\eta}{3-\kappa} (1 - \frac {\beta'_0}{(3-\kappa) 
{\mathcal D}}) \biggl [\frac {1}{{\tau - 1 + \frac{\eta}{3 - \kappa}}}(1 - 
\nonumber \\
& & \biggl (\frac {\beta'_{(0)}(r')}{\beta'_0} \biggr )^{\tau - 1 + 
\frac{\eta}{3-\kappa}}) + \quad \frac {(3-\frac{\tau}{2}) {\beta'_0}^2}
{\tau + 1 + \frac{\eta}{3 - \kappa}} (1 - \nonumber \\
& & \quad \biggl (\frac {\beta'_{(0)}(r')}{\beta'_0} \biggr )^{\tau + 1 + 
\frac{\eta}{3-\kappa}}) \biggr ] + \ldots \biggr \} \nonumber \\
\label {dynmzerofirst}
\eea
Similarly, for small $\eta$ we use (\ref{dynmzeroetazero}) and obtain:
\bea
& & {\mathcal M}_{(0)}(r') \approx \frac {\Delta r'_0}{(3-\kappa)} \biggl (
\frac{\gamma_0}{\beta'_0} \biggr )^{\tau} ((3-\kappa) {\mathcal D})^{1-
\frac{\eta}{3-\kappa}} {\beta'_0}^{\tau - 2 + \frac{\eta}{3-\kappa}} 
\nonumber \\
& & \quad \biggl \{\biggl [1 + \frac{(3-\frac{\tau}{2}) {\beta'_0}^2}{\tau + 
\frac{\eta}{3-\kappa}} + \frac{\eta}{3-\kappa}(1 - \frac{\beta'_0}{(3-\kappa)
{\mathcal D}})(1 + \nonumber \\
& & \quad \frac{(3-\frac{\tau}{2}) {\beta'_0}^2}{\tau + 1 + \frac{\eta}{3 - 
\kappa}})\biggr ] - \biggl (\frac {\beta'_{(0)}(r')}{\beta'_0} \biggr )^
{\tau - 2 + \frac{\eta}{3-\kappa}} \biggl [1 + \frac{(3 - \frac{\tau}{2}) 
{\beta'_{(0)}}^2(r')}{\tau + \frac{\eta}{3-\kappa}} + \nonumber \\
& & \quad \frac{\eta\beta'_{(0)}(r')}{(3-\kappa)\beta'_0}
(1 - \frac{\beta'_0}{(3-\kappa){\mathcal D}})(1 + \frac{(3-\frac{\tau}{2}) 
{\beta'_{(0)}}^2(r')}{\tau + 1 + \frac{\eta}{3-\kappa}})\biggr ] \biggr \} 
\nonumber \\
\label {dynmzerofirsteta}
\eea
For a quasi static active region and the same shock conditions as in 
(\ref{steadymzero}) the first order ${\mathcal M}_{(0)}(r)$ is:
\bea
& & {\mathcal M}_{(0)}(r') \approx \frac {\Delta r'_\infty}{(3-\kappa)} 
((3-\kappa) {\mathcal D})^{1-\frac{\eta}{3-\kappa}} (1 - \frac 
{\beta'_0}{(3-\kappa) {\mathcal D}}) \nonumber \\
& & \quad \biggl \{\frac{\eta}{3 - \kappa} \biggl [\frac {1}{\frac{\eta}
{3-\kappa} - 1}({\beta'_0}^{\frac{\eta}{3-\kappa} - 2} - 
\frac {{\beta'_{(0)}}^{\frac{\eta}{3-\kappa} - 1}(r')}{\beta'_0}) + 
\nonumber \\
& & \quad \frac {3}{\frac{\eta}{3-\kappa} + 1} 
({\beta'_0}^{\frac{\eta}{3-\kappa}} - \frac{{\beta'_{(0)}}^{\frac{\eta}
{3-\kappa} + 1}(r')}{\beta'_0})\biggr ] - \nonumber \\
& & \frac{(\eta + \delta)((3-\kappa) {\mathcal D})^{-\frac{\delta}
{3 - \kappa}}}{3 - \kappa} \biggl [\frac {1}{\frac{\eta + \delta}
{3-\kappa} - 1}({\beta'_0}^{\frac{\eta + \delta}{3-\kappa} - 2} - 
\frac {{\beta'_{(0)}}^{\frac{\eta + \delta}{3-\kappa} - 1}(r')}
{\beta'_0}) + \nonumber \\
& & \quad \frac {3}{\frac{\eta + \delta}{3-\kappa} + 1} ({\beta'_0}^
{\frac{\eta + \delta}{3-\kappa}} - \frac{{\beta'_{(0)}}^{\frac{\eta + \delta}
{3-\kappa} + 1}(r')}{\beta'_0})\biggr ] \biggr \} + \ldots 
\label {quasizerofirst}
\eea

\section {Spectrum and lags for power-law and exponential electron 
distributions} 
\label {app:b}
Using (\ref{nedistpow}) to (\ref{gammam}), after integration over $\gamma_e$ 
we obtain the following expression for the spectrum:
\bea
& & \frac {dP}{\omega d\omega} = \frac{\sqrt {3} e^2}{3\pi} r^2 \Delta r 
\frac{N_e}{\Gamma^4 (r)~(1-\beta (r))^3} \biggl \{2\int_{\gamma_m}^\infty 
d\gamma_e \biggl (\frac{\gamma_e}{\gamma_m} \biggr )^{-(p+1)} \nonumber \\
& & \quad \gamma_e^{-2} \int_{\frac{\omega'}{\omega'_c}}^{\infty} K_{5/3} 
(\zeta) d\zeta + {\mathcal G}(r) \biggl (\frac{11}{12\gamma_m}
\biggl [\frac{2^{-\frac{2}{3}}\biggl (\frac{\omega'}{\omega'_m} 
\biggr )^{-\frac{1}{3}}\Gamma (\frac{1}{3})}{\frac{p}{4} + 
\frac{1}{3}} \nonumber \\
& & \quad ~_1F_2(\frac{p}{4} + \frac{1}{3}; \frac{2}{3}, \frac{p}{4} + 
\frac{4}{3}; \biggl (\frac{\omega'}{2\omega'_m} \biggr )^2) + \nonumber \\
& & \quad\frac{2^{-\frac{4}{3}} \biggl (\frac{\omega'}{\omega'_m} 
\biggr )^{\frac{1}{3}}\Gamma (-\frac{1}{3})}{\frac{p}{4} + 
\frac{2}{3}}~_1F_2(\frac{p}{4} + \frac{2}{3}; \frac{4}{3}, \frac{p}{4} + 
\frac{5}{3}; \biggl (\frac{\omega'}{2\omega'_m} \biggr )^2)
\biggr ] - \nonumber \\ 
& & \quad \frac{21}{8\gamma_m^3} \biggl [\frac{2^{\frac{1}{3}} \biggl 
(\frac{\omega'}{\omega'_m} \biggr )^{-\frac{2}{3}}\Gamma (\frac{2}{3})}
{\frac{p}{4} + \frac{2}{3}}~_1F_2(\frac{p}{4} + \frac{2}{3}; \frac{1}{3}, 
\frac{p}{4} + \frac{5}{3}; \biggl (\frac{\omega'}{2\omega'_m} 
\biggr )^2) + \nonumber \\
& & \quad \frac{2^{-\frac{5}{3}} \biggl (\frac{\omega'}{\omega'_m} 
\biggr )^{\frac{2}{3}}\Gamma (-\frac{2}{3})}{\frac{p}{4} + \frac{4}{3}}~_1F_2
(\frac{p}{4} + \frac{4}{3}; \frac{5}{3}, \frac{p}{4} + \frac{7}{3}; \biggl 
(\frac{\omega'}{2\omega'_m} \biggr )^2)\biggr ] + \nonumber \\ 
& & \quad \frac{7 \gamma_m^2}{4} \biggl (\frac{\omega'}{\omega'_m}\biggr ) 
\int_{\gamma_m}^\infty d\gamma_e \biggl (\frac{\gamma_e}{\gamma_m} \biggr )^
{-(p+1)} \gamma_e^{-4} \int_{\frac{\omega'}{\omega'_c}}^{\infty} K_{5/3} 
(\zeta) d\zeta \biggr ) \biggr \} \nonumber \\
\label{powerpowlaw}
\eea
where $\omega'_m \equiv \omega'_{cc} \gamma_m^2$ is the minimum characteristic 
frequency of electrons. The double integrals in the first and last terms 
of (\ref{powerpowlaw}) do not have a simple analytical expression. An 
approximation can be obtained using an integral form of $K_\nu$ Bessel 
function:
\bea
& & K_\nu (\zeta) = \int_0^\infty dx e^{-\zeta ch~x} ch~\nu x 
\label {kbesseldef}\\
& & \int_\alpha^\infty K_\nu (\zeta) d\zeta = \int_0^\infty dx 
\frac {ch~\nu x}{ch~x} e^{-\alpha ch~x} = 2 K_{\nu-1} (\alpha) - \nonumber \\
& & \quad \int_0^\infty dx \frac{sh~(\nu - 1)x~sh~x}{ch~x}(\frac {1}{2ch~x} + 
\frac {3}{8ch~x} + \ldots) e^{-\alpha ch~x} \nonumber \\
\label{besselintappr}
\eea
The last integral is small and can be neglected. Therefore:
\bea
& & \int_{\gamma_m}^\infty d\gamma_e \biggl (\frac{\gamma_e}{\gamma_m} 
\biggr )^{-(p+1)}\gamma_e^{-2} \int_{\frac{\omega'}{\omega'_c}}^{\infty} 
K_{5/3} (\zeta) d\zeta \approx \nonumber \\
& & \quad \frac{1}{2\gamma_m}\biggl [\frac{2^{-\frac{1}{3}} \biggl 
(\frac{\omega'}{\omega'_m} \biggr )^{-\frac{2}{3}}\Gamma 
(\frac{2}{3})}{\frac{p}{4} + \frac{1}{6}}~_1F_2(\frac{p}{4} + \frac{1}{6}; 
\frac{1}{3}, \frac{p}{4} + \frac{7}{6}; \nonumber \\
& & \quad \biggl (\frac{\omega'}{2\omega'_m} \biggr )^2) + 
\frac{2^{-\frac{5}{3}} \biggl (\frac{\omega'}{\omega'_m} 
\biggr )^{\frac{2}{3}}\Gamma (-\frac{2}{3})}{\frac{p}{4} + 
\frac{5}{6}}~_1F_2(\frac{p}{4} + \frac{5}{6}; \frac{5}{3}, \nonumber \\
& & \quad \frac{p}{4} + \frac{11}{6}; \biggl (\frac{\omega'}{2\omega'_m} 
\biggr )^2)\biggr ] \label{firstkint} \\
& & \quad \int_{\gamma_m}^\infty d\gamma_e \biggl (\frac{\gamma_e}{\gamma_m} 
\biggr )^{-(p+1)}\gamma_e^{-4} \int_{\frac{\omega'}{\omega'_c}}^{\infty} 
K_{5/3} (\zeta) d\zeta \approx \nonumber \\
& & \quad \frac{1}{2\gamma_m^3}\biggl [\frac
{2^{-\frac{1}{3}} \biggl (\frac{\omega'}{\omega'_m} \biggr )^
{-\frac{2}{3}}\Gamma (\frac{2}{3})}{\frac{p}{4} + \frac{2}{3}}~_1F_2(\frac{p}
{4} + \frac{2}{3}; \frac{1}{3}, \frac{p}{4} + \nonumber \\
& & \quad \frac{5}{3}; \biggl (\frac{\omega'}{2\omega'_m} \biggr )^2) + 
\frac{2^{-\frac{5}{3}} \biggl (\frac{\omega'}{\omega'_m} 
\biggr )^{\frac{2}{3}}\Gamma (-\frac{2}{3})}{\frac{p}{4} + 
\frac{4}{3}}~_1F_2(\frac{p}{4} + \frac{4}{3}; \frac{5}{3}, \nonumber \\
& & \quad \frac{p}{4} + \frac{7}{3}; \biggl (\frac{\omega'}{2\omega'_m} 
\biggr )^2)\biggr ] \label{secondkint}
\eea
If the electron distribution is exponential i.e. $n'_e (\gamma_e) = N_e 
exp (-\alpha \gamma_e)$, terms with$~_1F_2$ must be replace by terms 
proportional to Mejier's G-functions. These functions do not have simple 
analytical presentations. Therefore only asymptotic behaviour of power 
spectrum is described in Sec. \ref{sec:synchflux}.

In the same way we can determine an analytical expression for lags when 
accelerated electrons have a power-law distribution using (\ref{kcoeff}) to 
(\ref{lagq}) and (\ref{powerpowlaw}) to (\ref{secondkint}). In fact, it is 
easy to see that the functions $P(r_0,\omega'_0)$ and $Q(r_0,\omega'_0)$ are 
both functionals of $\mathcal{H}$ which contains $\gamma_e$ dependent terms 
similar to the spectrum $dP/\omega d\omega$. The function $\mathcal{H}$ 
includes Bessel and hypergeometric functions and their derivatives. Therefore 
we can use (\ref{powerpowlaw}) to (\ref{secondkint}) to determine them.
As the derivatives of hypergeometric functions~$_1F_2 (\alpha, \beta, \gamma; 
z)$ are also hypergeometric, lags can be expressed as a sum of~$_1F_2$ 
functions. The calculation is straightforward but long and laborious and we 
do not present details here because their complexity does not permit to 
investigate their properties and numerical calculation is needed.

\section*{Acknowledgments}
I would like to thank Keith Mason for encouraging me to work on GRB science. 
The ideas presented in this work couldn't be developed without long 
discussions with the past and present members of the \swift science team at 
MSSL: A. Blustin, A. Breeveld, M. De Pasquale, P. Kuin, S. Oates, S. Rosen, 
M. Page, P. Schady, M. Still, and S. Zane, as well as the other members of 
the \swift team in particular: S. Barthelmy, Ph. Evens, E.E. Fenimore, 
N. Gehrels, P. M\'es\'zaros, J. Osborne, and K. Page. I thank all of them.

\end{document}